\documentclass[a4paper,11pt]{article}
\pdfoutput=1 

\usepackage{jcappub} 
\usepackage{graphicx}
\usepackage{subcaption}                 
\usepackage[T1]{fontenc} 
\usepackage{amsmath}

\definecolor{mycolor}{rgb}{1.0, 0., 1.0}

\title{\boldmath Dark matter capture by the Sun: revisiting velocity distribution uncertainties}

\author[a,b]{A. Nu\~nez-Casti\~neyra,}
\author[a]{E. Nezri,}
\author[b]{V. Bertin.}

\affiliation[a]{Aix Marseille Univ, CNRS, CNES, LAM, Marseille, France 
}
\affiliation[b]{Aix Marseille Univ, CNRS/IN2P3, CPPM, Marseille, France 
}

\emailAdd{arturo.nunez@lam.fr}
\emailAdd{emmanuel.nezri@lam.fr}
\emailAdd{bertin@cppm.in2p3.fr}

\abstract{
Among the different strategies aiming to detect WIMP dark matter (DM), a neutrino signal coming from the Sun would be a smoking gun. This possibility relies on the DM capture by the Sun driven by the local DM distribution assumptions: the local mass density and the velocity distribution. In this context, we revisit those astrophysical hypotheses (also relevant for direct detection). We focus especially on the DM velocity distribution considering different possibilities beyond the popular Maxwellian distribution. Namely, some alternatives can be considered through analytical approaches and by looking into cosmological simulations of spiral galaxies. 
Most of the fitting formulas used to constrain the local velocity distribution function fail to describe the peak and the high velocity tail of the velocity distribution  observed in simulations, the latter being improved when adding the local escape velocity of DM into the benchmark fitting models. In addition we test the predictions by the Eddington inversion method and also illustrate the importance of the galactic dynamical history. We estimate the resulting uncertainties on the DM capture rate by the Sun and conclude that different velocity distributions will affect the capture rate of DM by the Sun up to a $15-20\%$. On top of that, the calculation of the intrinsic variance of the capture rate leads to poorly controlled uncertainties especially for high WIMP masses ($>$30 GeV) raising concerns about the capture scenario.}

\begin{document}
\today
\maketitle

\flushbottom
\section{Introduction}\label{sec:intro}
While the search for particle dark matter (DM) remains far from solving the DM problem, enormous progress have been achieved in both direct and indirect detection efforts in the last decades. In particular,  with the weakly interacting massive particles (WIMPs) as one of the most popular DM candidates \cite{Jungman1996,primack1988detection,bergstrom2000reports}, its lack of direct or indirect detection shows the importance to understand the sources of uncertainties connected to the different experimental strategies. For direct or neutrino indirect detection the way the DM phase space is modeled impacts the derivation of experimental constraints and expectations of detectability (see e.g \cite{Choi2014,Green2011,2019JCAP...01..006F,Fairbairn:2008gz}). 

The WIMPs are good DM particle candidates within the $\Lambda$CDM scenario in many extensions beyond the Standard Model \cite{Martin:1997ns,appelquist2001bounds}. WIMPs are expected to have masses between some GeV up to a few TeV \cite{bertone2005particle} and to self-annihilate leading to fluxes of Standard Model particles. Neutrino fluxes with high energies coming from the Sun, for example, if detected in neutrino telescopes would be an unmistakable indirect detection of DM \cite{krauss1986solar,hagelin1986high,srednicki1987high, griest1987k}.
 Thus, neutrino telescope searches towards the Sun has become one of the main current strategies for DM detection, complementing direct detection to probe the local DM velocity distribution in the neighbourhood of the Sun. Namely, neutrino telescopes (resp. direct detection experiments) are essentially sensitive to the low (resp. high) velocity range. Consequently,  several studies have been performed to characterize  the prospects of detection of high energy neutrinos from the direction of the Sun \cite{bottino1991indirect,halzen1992f,bergstrom1998indirect,barger2002indirect,bertin2002neutrino,cirelli2005spectra,mena2008reconstructing,lehnert2008neutrino,barger2007high,liu2008neutrino,hooper2009new,kundu2012neutrinos}. Even if indirect detection with neutrino telescopes have failed to identify a signal, tight limits have been imposed on DM-nucleon interaction cross-section by different neutrino detectors \cite{adrian2016limits, choi2015search, aartsen2016improved,boliev2013search,demidov2015indirect}. The formalism and uncertainties related to DM capture by the Sun are treated in \cite{Jungman1996,Garani2017}, we revisit in this work the astrophysical uncertainties related to dark matter distributions such as the local DM density and velocity distribution. Different assumptions can be inferred from cosmological simulations, galactic dynamics and other approaches (see e.g \cite{Hansen:2005yj,Hansen:2008ek,2007PhRvD..75d3503V,Vergados:2014eya,2018arXiv180702519N,Cerdeno:2016znc}). Furthermore, DM only \cite{vogelsberger2009phase,Kuhlen:2009vh} or detailed  hydrodynamical simulations of Milky Way size halos inhabited by a disc galaxy often show departure from the standard Maxwellian description of the DM halo velocities distribution \cite{Ling2009,Kuhlen:2013tra,2016MNRAS.462..663B,Bozorgnia:2016ogo,Bozorgnia:2017brl,2018arXiv181012301N}.

  However direct extrapolation of simulated Milky Way size halos to experimental expectation calculation could bring inconsistencies.  Resolution issues or deeper differences between the nature of the simulated galaxy and the real one make it impossible to guarantee compatibility between both galaxies. Nevertheless, cosmological simulations can still be used to validate theoretical approaches aiming to predict the phase space distribution of a self gravitating object from its gravitational potential \cite{Lacroix2018,pascale2018action,posti2019mass,LacroixNunez}.

  In this paper, we study the astrophysical assumptions relevant to the DM capture by the Sun focusing on the velocity distribution. Specifically, we estimate the related uncertainties by using analytical methods and cosmological simulations.
  The paper is organized as follows. In section \ref{sec:formalism}, we recall the mechanism of stellar DM capture and its intrinsic uncertainties. In section \ref{sec:ansatz}, some of the most common velocity distribution functions are described and compared in section \ref{sec:simus} with the velocity distributions in simulations. We compare the velocity distribution simulation outcome with the Eddington inversion method predictions in section \ref{sec:Dynamics}. In section  \ref{sec:consequences} we show the subsequent effects on DM capture induced by the different assumptions. Finally, the results are summarized in section \ref{sec:Summary}.

\section{Dark matter capture by the Sun}\label{sec:formalism}
\subsection{Preamble and notations}
The description and formalism of WIMP capture by the Sun can be found in e.g \cite{Gould1987,griest1987k,Gould1990,Gould1992,Jungman1996,Garani2017}. The change in the number, $N$, of captured WIMPs inside the Sun is described by the competition between the rate of capture, $C$, and  two times the annihilation rate of the already captured WIMPs in the centre of the Sun, $\Gamma_A = \frac{1}{2} C_A N^2$ (as long as the DM mass $m_{\chi} > 5\;\rm GeV$ where WIMP evaporation is not relevant \cite{griest1987k,Gould1987,Garani2017,Busoni2017}). Once the age of the Sun is greater than the time needed for the capture and annihilation processes to equilibrate, $\tau_{\odot} = (CC_A)^{-1/2}$, the annihilation rate becomes proportional to one half of the capture rate. Then, considering that the age of the Sun, $t_{\odot}$, is around $4.5$ Gyr and following \citep{Jungman1996,griest1987k}  we can write the age of the Sun in terms of the equilibrium time scale as:
\begin{equation}\label{eq:equilibrium}
    \frac{t_{\odot}}{\tau_{\odot}}= 330 \left( \frac{C}{s^{-1}}\right)^{1/2} \left( \frac{\langle \sigma_A v\rangle}{\textrm{cm$^3$ s$^{-1}$}} \right)^{1/2} \left(\frac{m_{\chi}}{10\; \textrm{GeV}} \right)^{3/4}
\end{equation}
where $C$ is the capture rate in the Sun, $\langle \sigma_A v \rangle$ is the total annihilation cross section times the relative velocity in the limit where $v \to 0$ and $m_{\chi}$ is the WIMP mass. Once in equilibrium, $t_{\odot}/\tau_{\odot}>1$, the capture rate process becomes the main process to describe the expected neutrino flux from DM annihilation inside the Sun. The differential capture rate per unit shell volume by a nucleus $i$ at a distance $r$ from the centre of the Sun is:
\begin{equation}\label{eqCapture}
    \frac{dC_{i}}{dV}=\frac{\rho_{\odot}}{m_{\chi}} \int_0^{u^{\odot}_{esc}} du \frac{f(u)}{u} \Omega_i(Q)
\end{equation}

\noindent where $\rho_{\odot}$ is the dark matter density in the solar neighbourhood. As stated here, the capture consists in the convolution of the local DM velocity distribution function (VDF) \footnote{As commonly done in the literature, in this paper we use abusively {\it velocity distribution} instead of {\it speed  distribution}.} expressed in the frame of reference of the Sun, $f(u)$, where the maximal speed is defined as $u^{\odot}_{esc} = v_{esc} + v_{\odot}$ with $v_{esc}=v_{esc}(r=r_{\odot})$ the escape velocity from the galactic potential at the solar radius and $v_{\odot}$ the circular velocity of the Sun around the galactic centre. $\Omega_i$ is the probability that a WIMP traveling at a speed $u$ ends up with a speed smaller than the solar escape speed at a radius $r$ from the Sun centre, $u_{esc}(r)$, after being scattered from a nucleus $i$ inside the Sun, rendering the WIMP a captured WIMP \cite{Silk1985}. This probability is modeled as:

\begin{equation}
    \Omega_i =\sigma_i n_i \frac{M_i}{2 \mu_i^2} \int_{Q_{min}} ^{Q_{max}} F_i^2 (Q) dQ
\end{equation}
\begin{equation*}
    Q_{min} = \frac{1}{2} m_{\chi} u^2 \;\;,\;\; Q_{max}=\frac{1}{2}\beta_{+} m_{\chi} w^2 \;\;,\;\; \beta_{+}=\frac{4 M_i m_{\chi} }{(M_i + m_{\chi})^2}\;\;,\;\; w^2 = u^2 + u^2_{esc}(r)
\end{equation*}
\noindent with $F_i^2(Q)$ being an exponential (Helm) nuclear form factor \cite{Gould1987,Helm1956}, $\sigma_i$ the WIMP-nucleus elastic scattering cross-section in the limit of zero momentum transfer, $n_i$ and $M_i$ the number density and the mass of the nucleus respectively, $\mu_i$ the WIMP-nucleus reduced mass and $Q$ is the recoil energy.

In the case of the spin dependent (SD) interactions only scattering on hydrogen is considered, and for spin independent (SI) case isospin conserving scatter is assumed. The zero momentum transfer cross section on the nucleus is $\sigma_i^{SD} = \sigma_p^{SI} I^2 \frac{\mu_i ^2}{\mu_p ^2}$ where $I$ and $\mu_p$ are the mass number of the nucleus and the WIMP-nucleon reduced mass respectively.

The final capture rate is described as follows:
\begin{equation}
C = \int_0 ^{R_{\odot}}4 \pi r^2 dr \sum_i \frac{dC_i}{dV} \;\; . 
\end{equation}

The VDF $f(u)$ in the frame of reference of a star moving around the galactic centre with a circular velocity $\vec{v}_{\odot}$ can be computed from the VDF in the galactic frame $f(v)$ through 
\begin{equation}
    f(u) = \int_{-1}^1 f{\left(\sqrt{v^2+v_{\odot}^2+2vv_{\odot}\cos{\theta}} \right)}d\cos{\theta}
\end{equation}

\noindent where $\vec{u} = \vec{v} + \vec{v}_{\odot}$  and $\theta$ is the angle between $\vec{v}$ and $\vec{v}_{\odot}$.

\begin{figure}[tbp]

\centering 
\begin{subfigure}[b]{0.47\linewidth}
    \includegraphics[width=\linewidth]{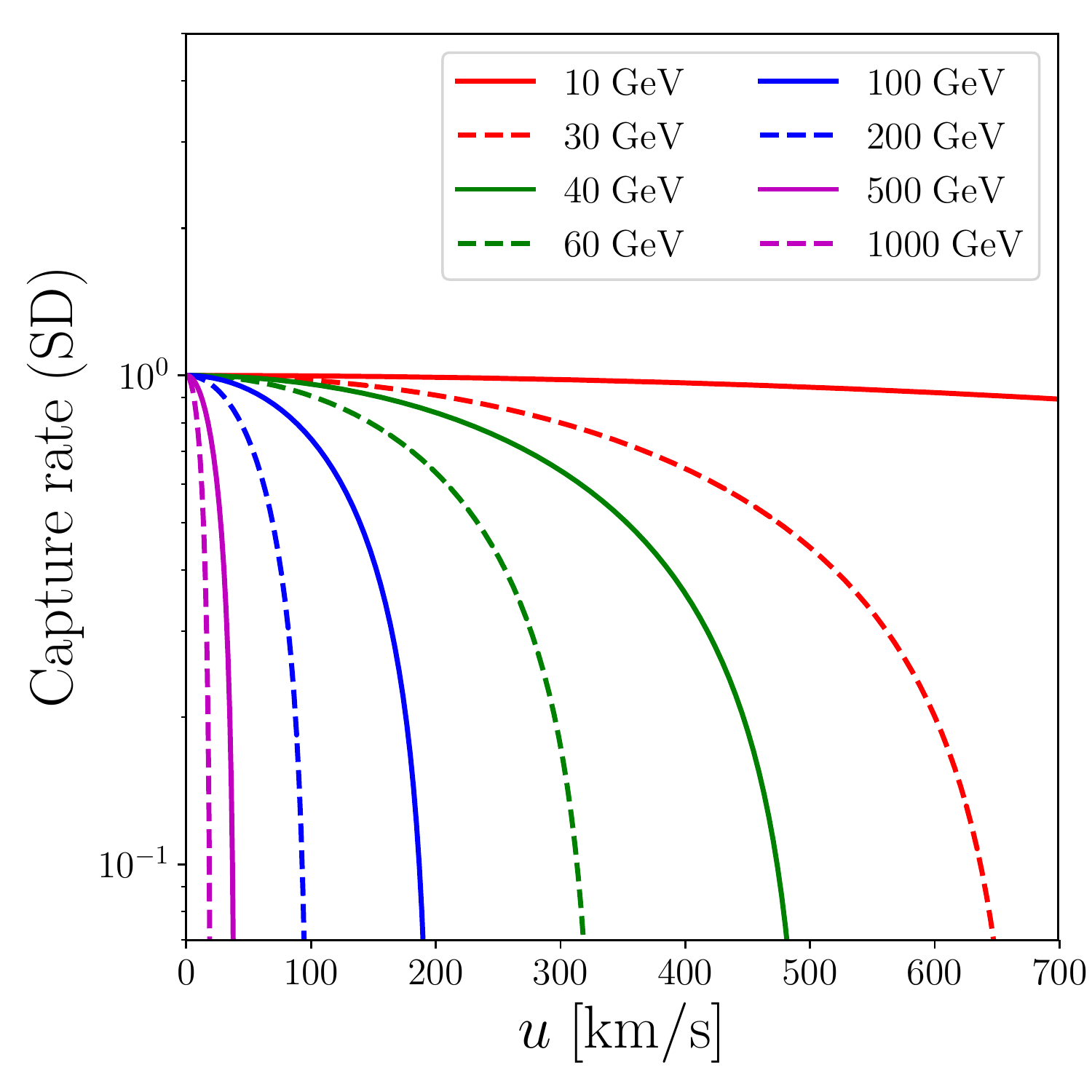}
    \caption{}
  \end{subfigure}
  \begin{subfigure}[b]{0.47\linewidth}
    \includegraphics[width=\linewidth]{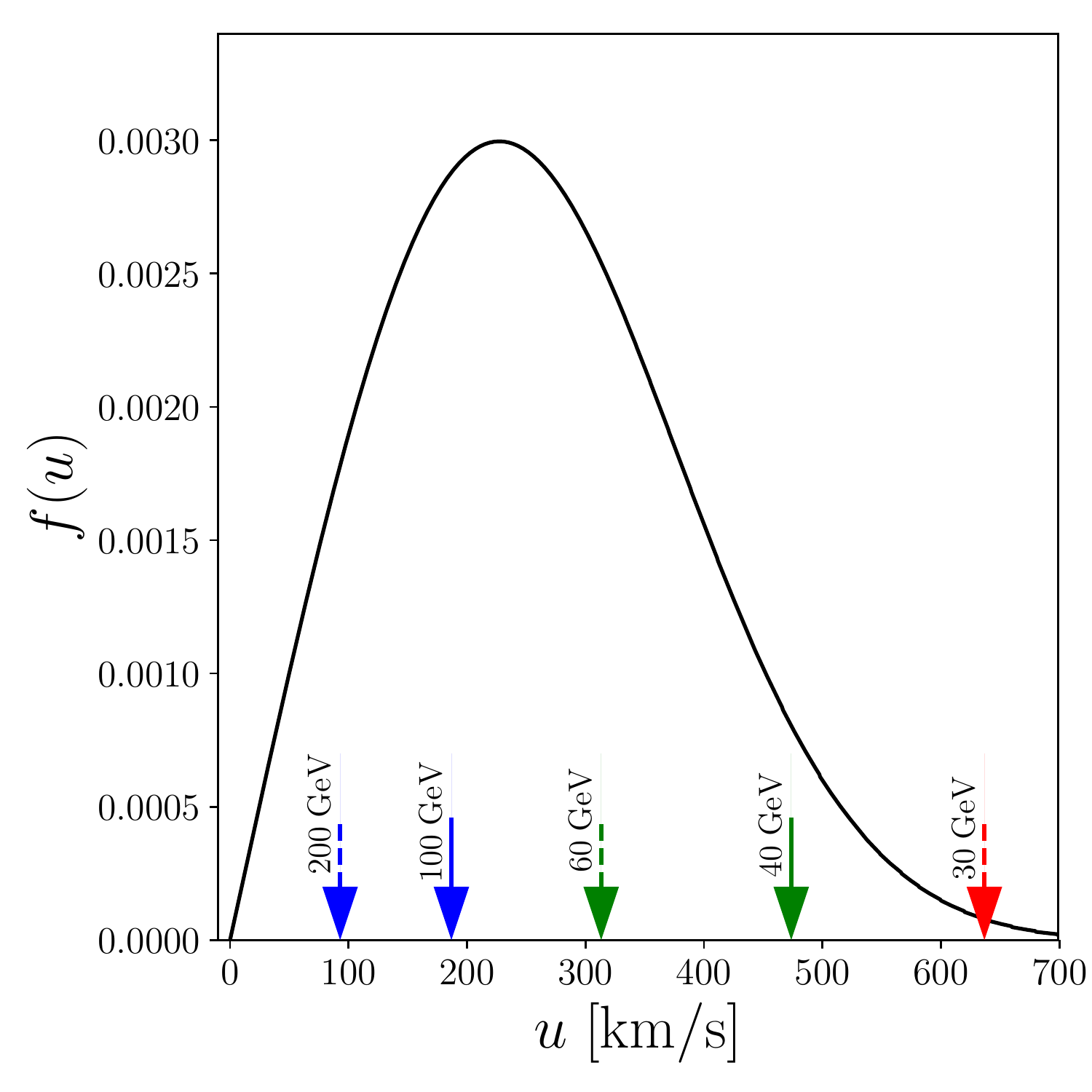}
    \caption{}
  \end{subfigure}
  \label{fig:coffee}
\caption{\label{fig:PP} (a) Dark matter capture rate computed setting the velocity distribution to a delta function in equation \ref{eqCapture} normalized to its maximum value for different dark matter masses ranging from $5\;\rm GeV$ up to $10 \;\rm TeV$ for the spin dependent case. (b) Maxwellian velocity distribution in the frame of reference of the Sun. The maximal velocity for which the capture is possible for different $m_{\chi}$, are shown as vertical arrows keeping the color notation of the left panel.}
\end{figure}

The probability for a DM particle to get captured is directly related to its ability to transfer enough recoil energy $Q$ to the nucleus during the interaction. If the transferred energy is smaller than $Q_{min}$, the WIMP is able to escape from the Sun. Figure \ref{fig:PP} shows the capture rate calculated using a delta function as the velocity distribution and normalized to their maximum value. As can be seen by comparing the two panels of figure  \ref{fig:PP}, heavy WIMPs are captured only when they populate the low velocity tail of the VDF, reducing the WIMP population concerned by the capture, and therefore the final neutrino flux as $m_{\chi}$ increases. The opposite effect is seen for lightest candidates where the energy transfer is efficient enough to yield a high probability of capture even for  WIMPs with velocities close to $v_{esc}+v_{\odot}$. The neutrino signal coming from low mass WIMPs subject to evaporation ($m_{\chi}>5$ GeV) in the Sun falls below the energy threshold of neutrino telescopes such as ANTARES, KM3NeT and IceCube. 

\subsection{Standard Halo Model}
 
The most popular set of benchmark assumptions describing the features of the Milky Way dark matter halo is called the Standard Halo Model (SHM).  It is the default model used in the field of local dark matter detection: direct detection and DM searches with neutrino telescopes. It consists on an isotropic, isothermal sphere DM halo with a density profile  $\rho \propto r^{-2}$ that yields the so-called Maxwellian distribution as the natural solution of the collision-less Boltzmann equation \cite{Green2011,Ullio2000}  having the following form:
\begin{equation}
    f(\vec{v}) = \frac{N}{2\pi v_0^2} \exp{\left( -\frac{3|\vec{v}|^2}{2 v_0^2} \right)}
\end{equation}
where the dispersion, $v_0$, is related to the circular velocity at 8 kpc, $v_c$,  as $v_0 = \sqrt{3/2}v_c$. The Maxwellian distribution formally extends up to infinity, so to avoid having a non-zero population  above the local escape velocity $v_{esc}$, an exponential cutoff is usually assumed. The popularity of this approximation resides on its simplicity. Its potential ability of reproducing what is observed in dynamical systems is discussed in section \ref{sec:simus}. The SHM also includes other astrophysical values (see e.g \cite{Green2011}), such as, the local dark matter density $\rho_{\odot} = 0.3 $ GeV/cm$^3$ = 0.01 $M_{\odot}$/pc$^3$, $v_{\odot} = 220 $ km/s, and the galactic escape velocity at the solar radius, $v_{esc}= 544$ km/s.


\subsection{Looking at the variance of the process: reliable capture rate estimation ?}\label{sec:uncertainties}

Equation \ref{eqCapture}
can be regarded as the expectation value of a function $\Omega_i(Q)/(u m_{\chi})$ with a probability distribution $f(u)$, therefore the associated variance can be computed as:

\begin{equation}\label{eq:variance}
\mathrm{Var}\left[\frac{dC}{dV}\rho_{\odot}^{-1}\right]= \sigma_C^2 = \int_0^{u_{esc}} f(u) \left(\frac{\Omega_i(Q)}{u m_{\chi}} -  \frac{dC}{dV}\rho_{\odot}^{-1}\right)^2 du \;\;\; .
\end{equation}

\begin{figure}[tbp]

\centering 
\begin{subfigure}[b]{0.47\linewidth}
    \includegraphics[width=\linewidth]{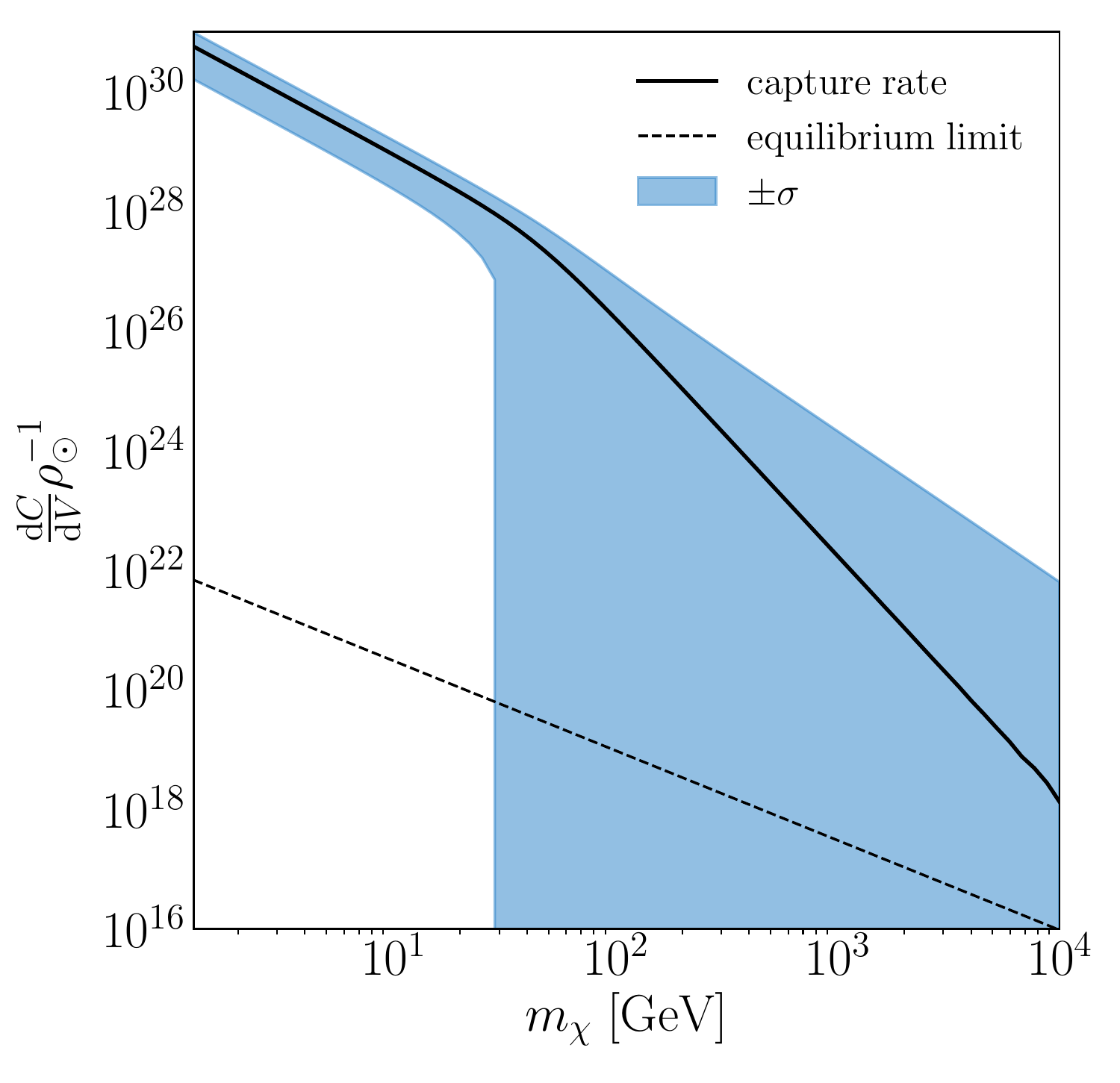}
    
    \caption{ Capture rate}
  \end{subfigure}
  \begin{subfigure}[b]{0.47\linewidth}
    \includegraphics[width=\linewidth]{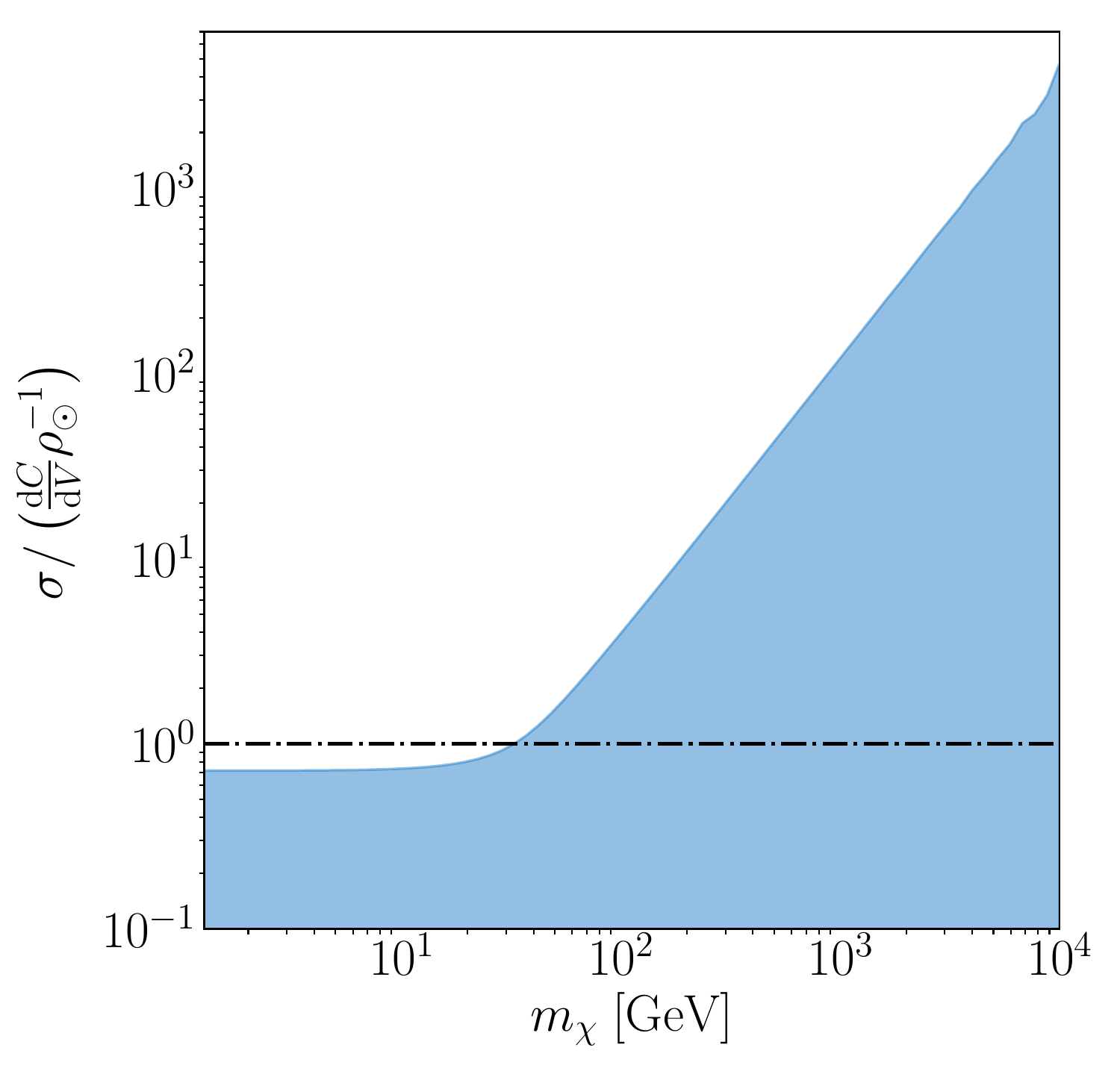}
    
    \caption{ Relative uncertainty on capture}
  \end{subfigure}

\caption{  \label{fig:SHM_uncertainty} (a) Differential capture rate per unit volume in the Sun considering the Standard Halo Model (black solid) with the one sigma band uncertainty (blue). The dashed line shows the equilibrium limit where the age of the Sun equates the equilibrium time scales. (b) The relative uncertainty of the capture process is shown by the blue band. The horizontal line shows the value where the relative uncertainty reaches $100\%$ of the dark matter capture. }
\end{figure}

As $m_{\chi}$ increases the probability $\Omega_i(Q)$ remains non-zero 
only in a restricted range of values of $u$ in the low velocity part of the VDF. This cut in the velocity distribution starts around  $m_{\chi}\gtrsim 30$ GeV for the relevant velocity range over which $f(v)$ spreads (see figure \ref{fig:PP}). Then the differential capture rate decreases, becoming up to 2-3 orders of magnitude smaller than the probability $\Omega_i(Q)$ alone for very high $m_{\chi}$. This increasing difference causes the uncertainty bands shown to grow as heavier DM particles are considered (figure \ref{fig:SHM_uncertainty}). For high WIMP masses the captured rate could thus fluctuate within several orders of magnitude, rising the concern of whether capture and annihilation inside the Sun could fall out of equilibrium. The value at which equation \ref{eq:equilibrium} is equal to 1 for a particular $m_{\chi}$ (using the thermal value $\langle \sigma_A v \rangle=3\times10^{-26} \rm{cm^3/s}$) \textit{i.e.} the limit below which capture and annihilation are no longer in equilibrium inside the Sun is also shown in figure \ref{fig:SHM_uncertainty}. We thus argue that predictions made for WIMPs heavier than 30 GeV need to be treated with strong caution given the high uncertainties ($\gg 100 \%$) related to their capture and the capture rate initial assumptions.

\section{Local dark matter distribution}\label{sec:DM_dist}

The dark matter distribution near the Sun is extensively reviewed and discussed in \cite{Read:2014qva} (see also the discussion in \cite{Famaey:2015bba}).
Different DM VDF assumptions can reduce or enhance the available population of WIMPs that can be captured, consequently affecting the sensitivities of DM detection efforts \cite{Choi2014,Ling2009,Green2011,Fornengo2017,Mao2013}. Concerning the DM VDF in the solar neighbourhood, several distribution functions have been proposed as attempts to better reproduce some features of the VDF shape, such as the platykurtic nature of the peak or the sharp cut of the tail of the distributions observed in cosmological simulation data. One of the main approaches to this problem is directly fitting ansatz to the data from cosmological simulations \cite{Ling2009, guedes2011forming, vogelsberger2009phase, bozorgnia2016simulated, Choi2014, Mao2013}. In this section we review some of the popular distribution functions proposed to fit the velocity distributions observed in dynamical systems.

\subsection{Analytical functions}\label{sec:ansatz}
Some of the properties observed in VDF obtained from cosmological simulations are hard to reproduce with classical Gaussian or Maxwellian functions \cite{Ling2009}. Those discrepancies can be reduced by introducing the generalization of these two functions. The generalized Maxwellian distribution is defined as:

\begin{equation}\label{eq:GenMaxwellian}
f(\vec{v}) = \frac{1}{N(v_0,\alpha)} e^{-(\vec{v}^2/v_0^2)^{\alpha}} \ .
\end{equation}
The normalization factor is $N(v_0,\alpha) = 4\pi v_0^3 \Gamma(1+3/2\alpha)$. The original Maxwellian distribution can be recovered by setting $\alpha=1$. In the same way a generalized Gaussian distribution can be defined. An alternative approach is  to derive an empirical distribution fitted on a set of DM halos of different sizes \cite{Mao2013}. This distribution has the particularity that for the low velocity tail it approaches an exponential distribution instead of a Gaussian distribution.

It has been pointed out that systems formed due to long range gravitational forces, such as the DM structures, should be treated by non extensive statistical mechanics \cite{Ling2009}. With the previous distribution functions it is hard to reproduce the flattened trend of the distributions frequently observed in simulations as well as the sharp drop of the high velocity tail without explicitly imposing so. The sharp drop observed in simulations is natural to relaxed collision-less structures \cite{Hansen2005}. A distribution function derived from a generalization of the Boltzmann-Gibbs approach by Tsallis has been proposed as a more appropriate function to fit the outcome of simulations, this functions has the following form:

\begin{equation}\label{eq:Tsallis}
f(\vec{v}) = \frac{1}{N(v_0,q)} \left(1-(1-q)\frac{\vec{v}^2}{v^2 _0}\right)^{q/(1-q)}
\end{equation}

\noindent where $N(v_0,q)$ is the normalization factor. For this distribution function the Maxwell-Boltzmann distribution is recovered when $q\to 1$. Numerical and analytical procedures have shown that self-gravitating collision-less structures in dynamical equilibrium follow Tsallis distributions \cite{Lima:2002hm,Hansen2005,Hansen:2004dg}. Within the Tsallis distribution a finite escape velocity is established as  $v_{esc}=v_0^2/(1-q)$ while in the Maxwellian and Gaussian distributions, the $v_{esc}$ is formally infinite. Therefore the $q$ parameter can be recovered from a dynamical construction of $v_{esc}$ as shown in the next section.

\subsection{Hints from simulations}\label{sec:simus}

\begin{figure}[h!]
  \centering
  \begin{subfigure}[b]{0.47\linewidth}
    \includegraphics[width=\linewidth]{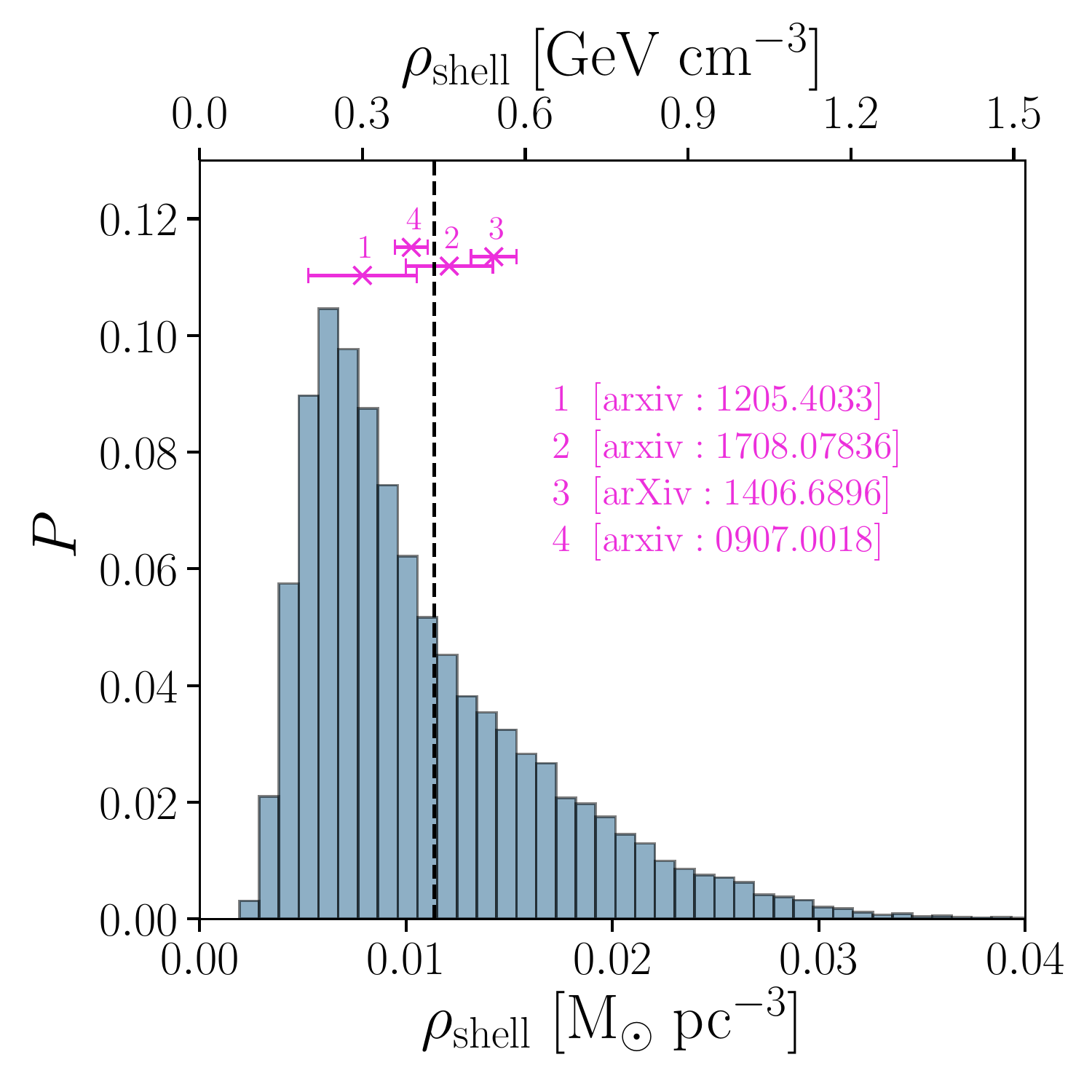}
    \caption{Halo B shell}
  \end{subfigure}
  \begin{subfigure}[b]{0.47\linewidth}
    \includegraphics[width=\linewidth]{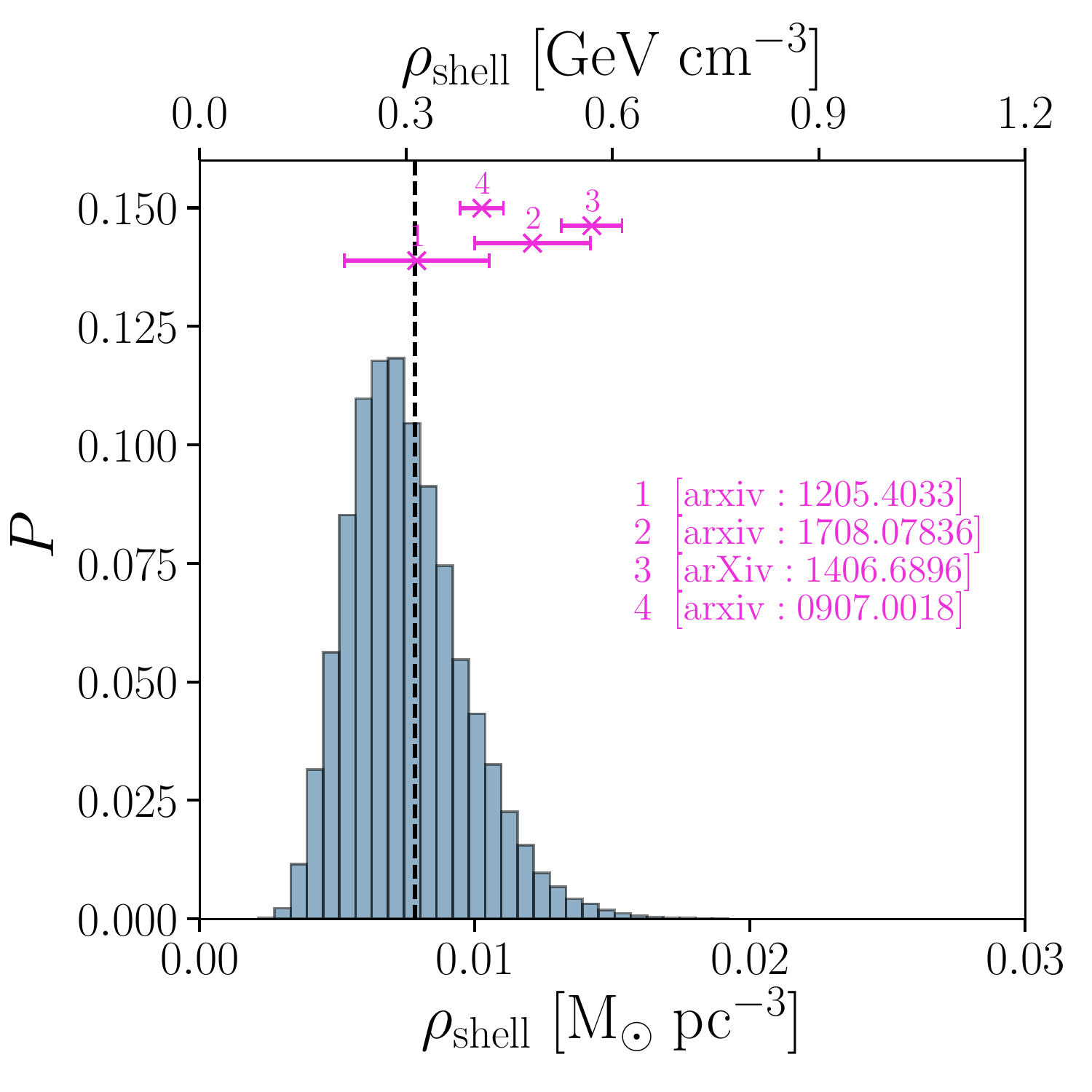}
    \caption{Mochima shell}
  \end{subfigure}
  \label{fig:coffee}
    \begin{subfigure}[b]{0.47\linewidth}
       \includegraphics[width=\linewidth]{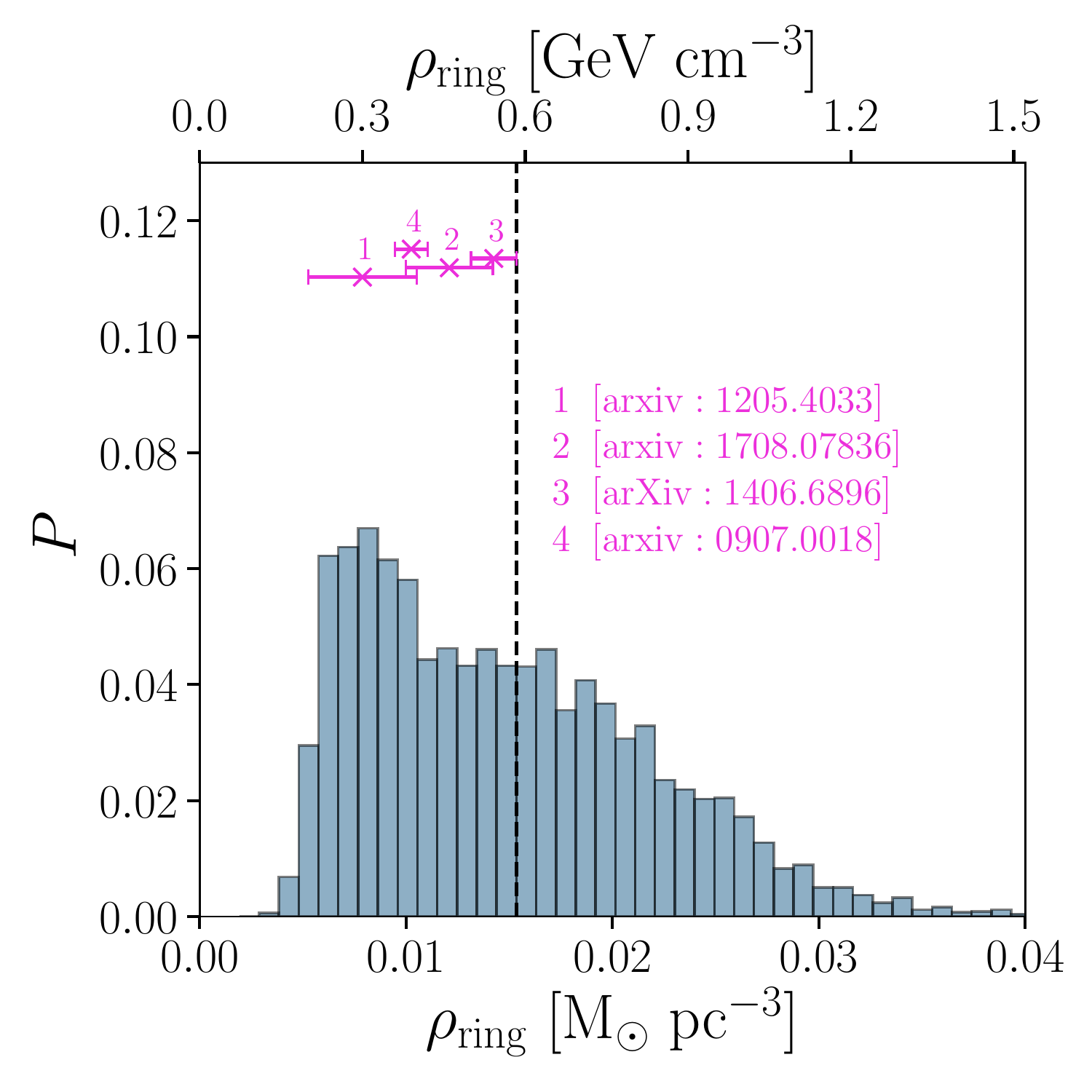}
        \caption{Halo B ring}
    \end{subfigure}
  \begin{subfigure}[b]{0.47\linewidth}
    \includegraphics[width=\linewidth]{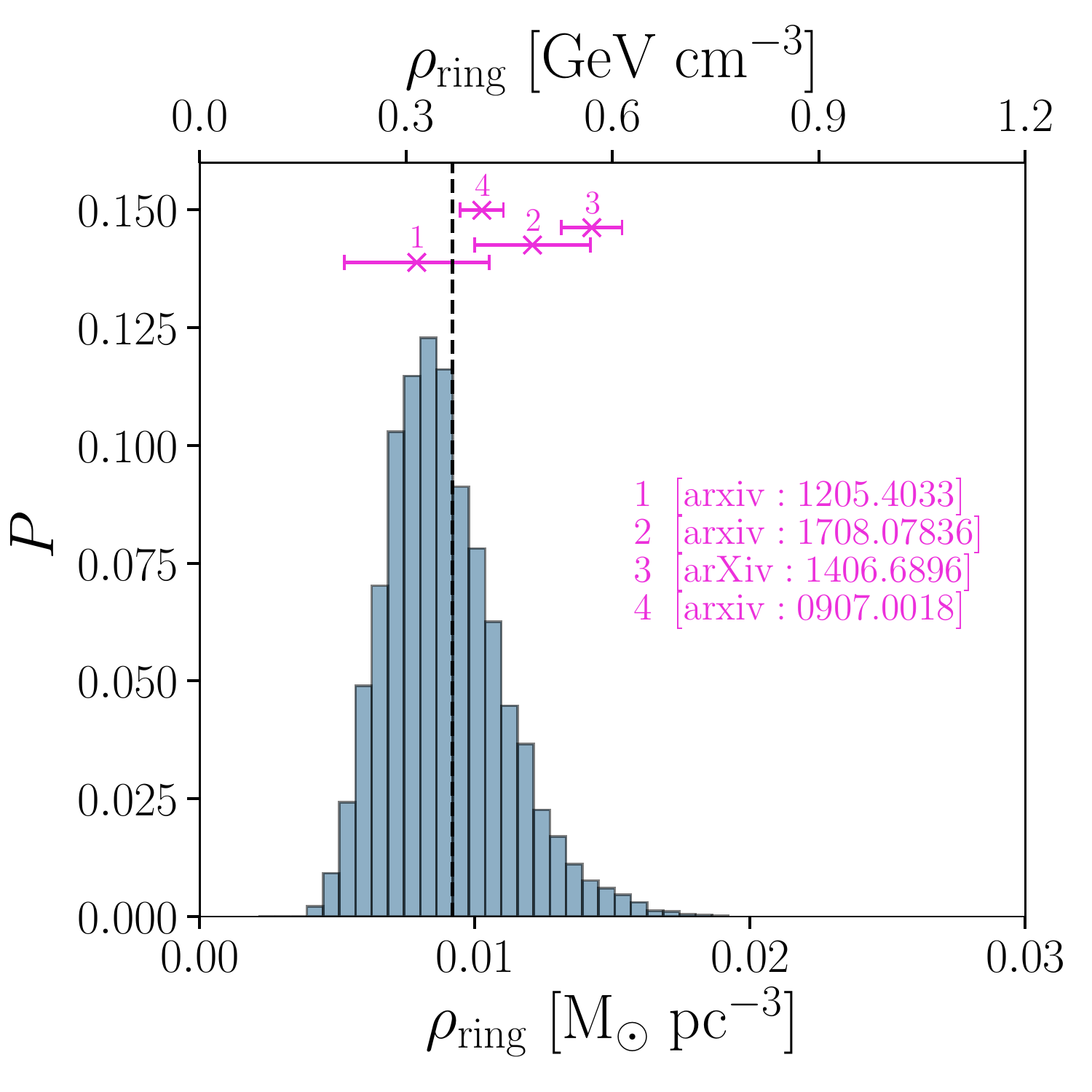}
    \caption{Mochima ring}
  \end{subfigure}
  \caption{Distribution of the DM particle density in $7\;$kpc $< r < 9\;$kpc from the galactic centre inside a shell (upper panels) or a square torus set in the galactic plane (bottom panels) for both simulations. Measurements with their error bars are also shown.}
  \label{fig:rho}
\end{figure}

The level of resolution reached in cosmological hydrodynamical simulation of ``Milky Way like'' \footnote{We emphasize here that the label ``cosmological simulations of Milky Way like galaxy'' is (abusively) used in the literature for ``cosmological simulations of (more or less) spiral galaxy embedded in a Milky Way size halo'' with poor checking of the merger and star formation history, the stellar-to-halo mass ratio, the morphology, etc... } galaxies allow such simulations to be a tool in the understanding of DM distribution in regions targeted by detection experiments. Of course the uncertainties cannot be bracketed by a single or few numerical objects. Due to the lack of statistics of such high resolution simulations, we use two state of the art simulations as illustrating examples. Given the limited resolution, we consider those simulations as diagnostic tools to test and calibrate analytical methods as well as insights for the shape of the velocity distribution function of the dark matter. As realistic Milky Way is out of reach, direct extrapolation might be considered as a biased approach. Here we examine the  evolution of the density and velocity of DM at $\sim$8 kpc from the centre of the halo in simulations of spiral galaxies inside a cosmological environment.

For this work we use two ``Milky Way like'' simulations performed with the RAMSES  \cite{Teyssier2002} and MUSIC \cite{hahn2011multi} packages using the zoom-in technique and including star formation and supernovae feedback implementations. The first one has been published in \cite{Mollitor2014} and the second simulation is a higher resolution simulation that forms part of a suite of simulations, to be described in a forthcoming
paper, where different sub-grid
physics for star formation and stellar feedback are considered \cite{Nunezsimu}. Both simulations are labeled Halo B and Mochima respectively.
 
In order to establish the validity of the simulations for the study presented in this paper, we check the local DM density between 7 and 9 kpc from the centre of the gravitational potential of the galaxy. We take two volumetric selections, both centered in the galactic centre: 1) the ring: a torus aligned with the galactic plane, with inner and outer radius of 7 kpc and 9 kpc respectively and a height of 2 kpc; 2) the shell: a spherical shell  with the same inner and outer radius that includes the ring selection within itself. We found that the spread and the mean of the sample of density measurements is consistent with the experimental results on the real Milky Way (see e.g \cite{Catena:2009mf,Bovy:2012tw,Bienayme:2014kva,Sivertsson:2017rkp,Read:2014qva,Famaey:2015bba}) as shown in Figure \ref{fig:rho}.
 In the following, we look at the evolution of the local density and velocity distribution of DM in the solar neighbourhood to asses whether the assumption of equilibrium between the capture and annihilation processes inside the Sun holds in the considered cosmological simulations. We explore some of the challenges of fitting velocity distributions in simulations and the impact of their use in calculations of the DM capture by the Sun.
\subsubsection{Time evolution}\label{sec:timeev}

\begin{figure}[h!]
  \centering
  \begin{subfigure}[b]{0.47\linewidth}
    \includegraphics[width=\linewidth]{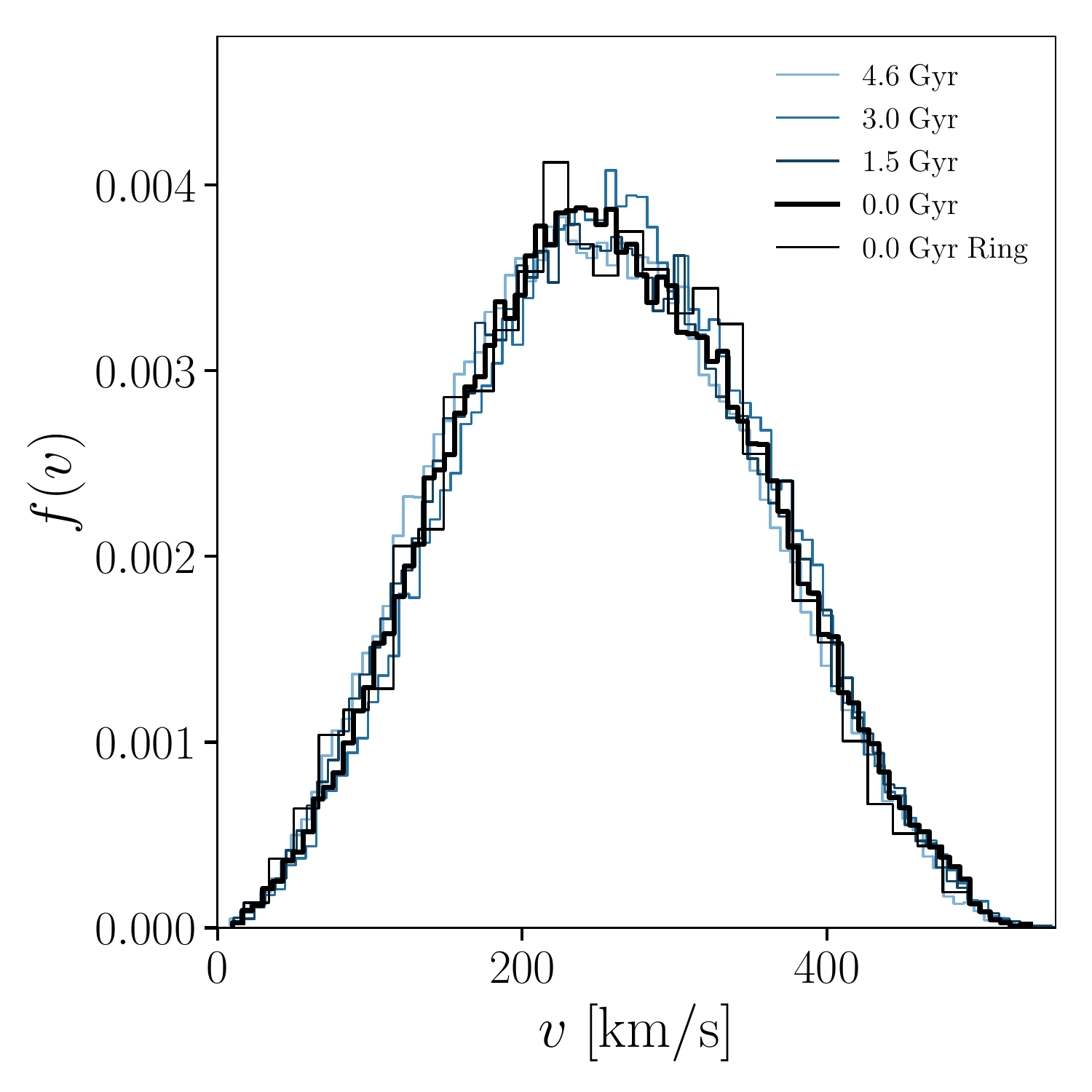}
    \caption{Halo B shell.}
  \end{subfigure}
  \begin{subfigure}[b]{0.47\linewidth}
    \includegraphics[width=\linewidth]{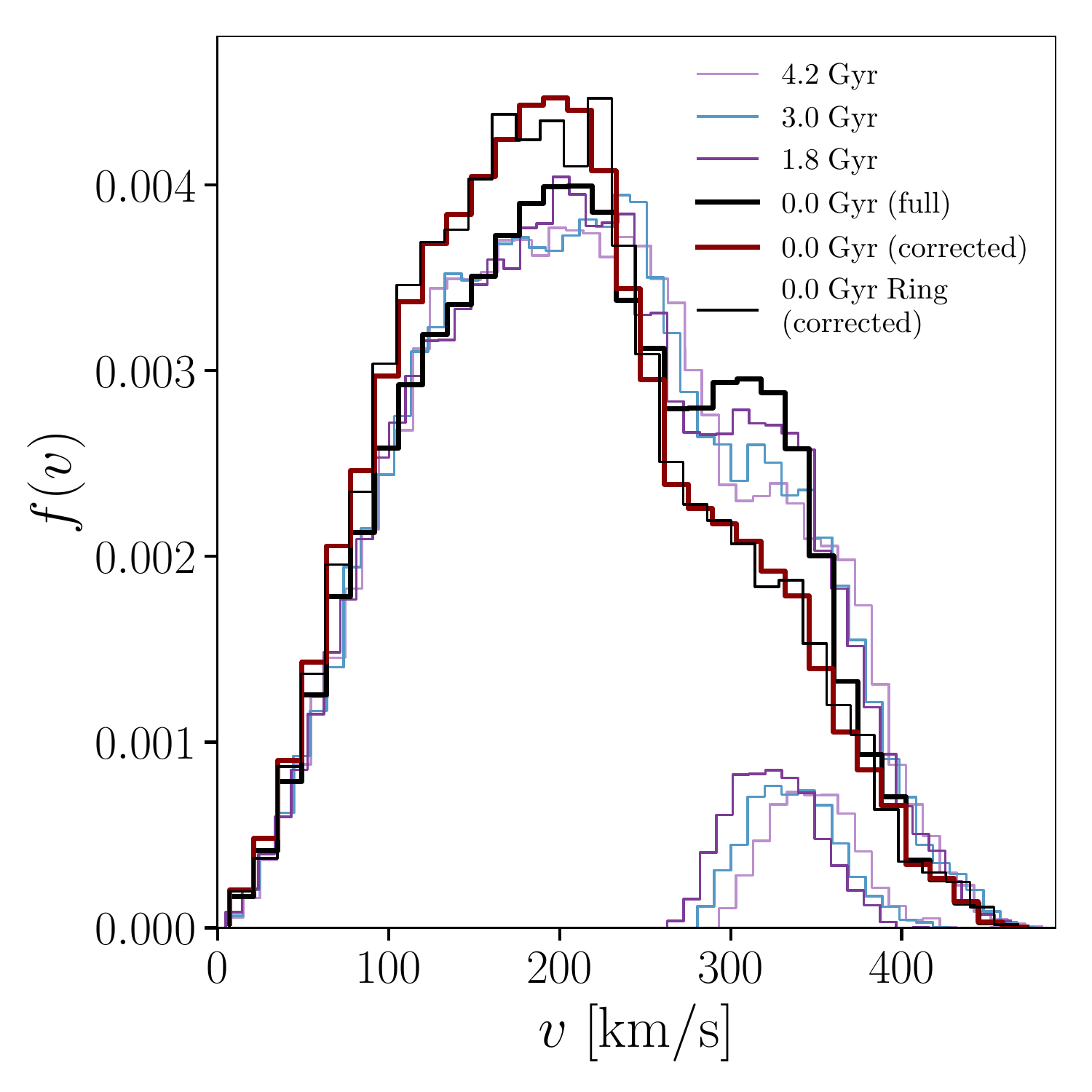}
    \caption{Mochima shell}
  \end{subfigure}
  \label{fig:coffee}
    \begin{subfigure}[b]{0.47\linewidth}
       \includegraphics[width=\linewidth]{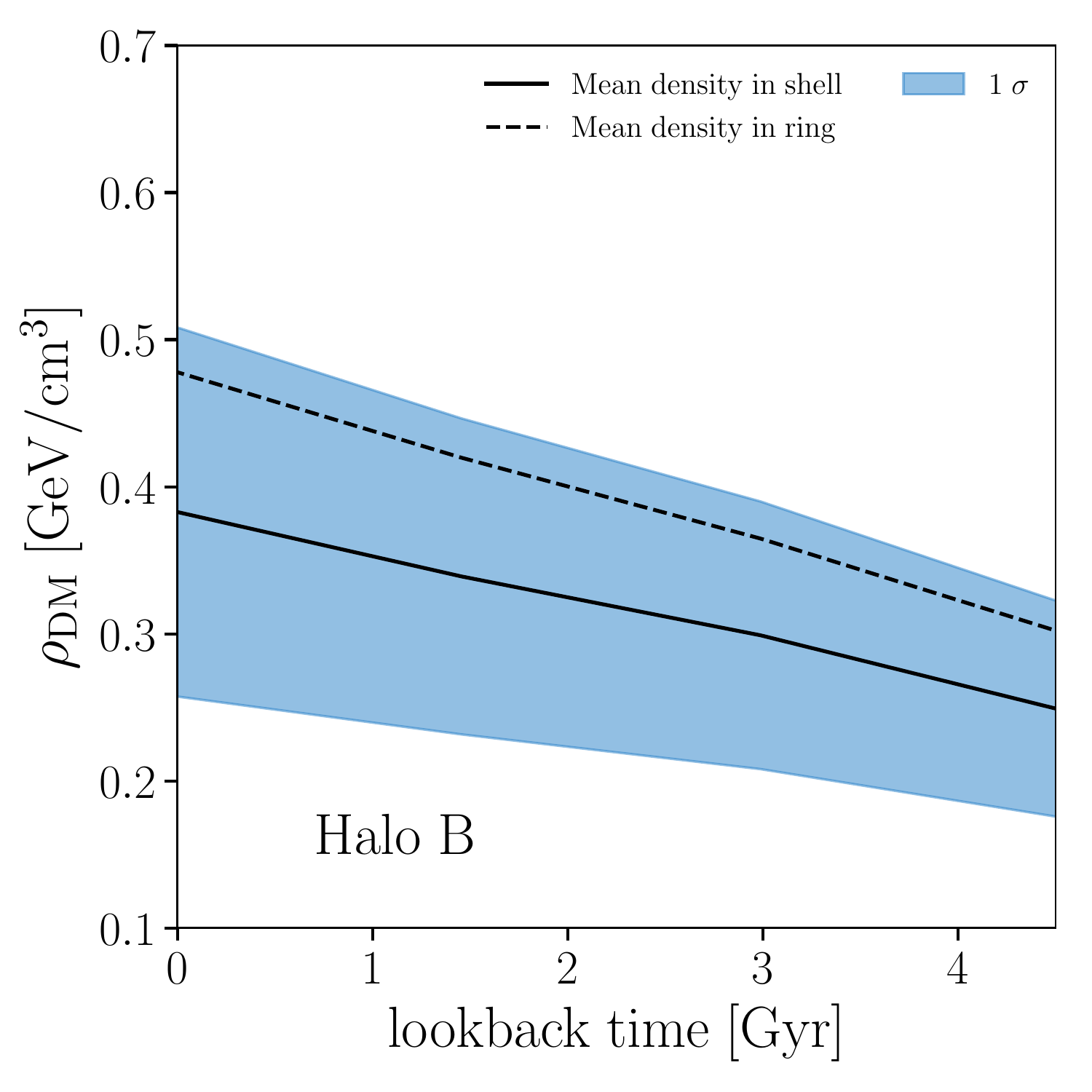}
        \caption{Halo B}
    \end{subfigure}
  \begin{subfigure}[b]{0.47\linewidth}
    \includegraphics[width=\linewidth]{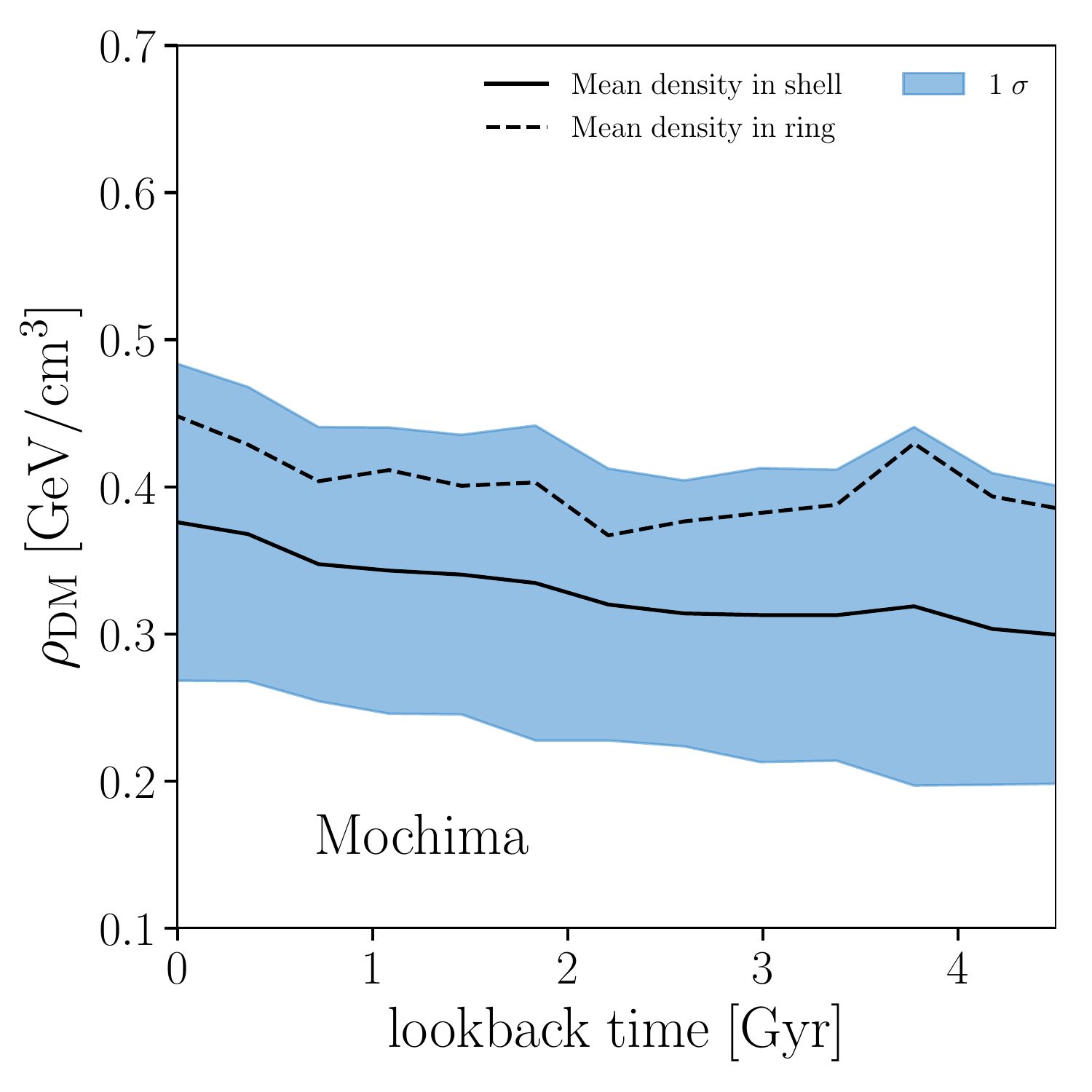}
    \caption{Mochima}	
  \end{subfigure}
  \caption{Time evolution of the DM velocity distribution (top panels) and of the mean density (bottom panels) at a radius of 8 kpc from the galactic centre for each simulation. In the case of Mochima, we show the full distribution, the corrected one and the removed population (see text).}
  \label{fig:timeEvolution}
\end{figure}

Equation \ref{eq:equilibrium} relies on assuming that both the local density and the local velocity distribution of DM have been stable during the lifetime of the Sun. The stability of those two quantities is closely related to the merging history of the galaxy. Recent works using {\it Gaia} DR2 results report that the Milky Way (MW) might not be in equilibrium having been perturbed by a crossing of the Sagittarius cloud through the galactic  disc $~300-900$ Million years ago \cite{antoja2018dynamically}. Such evidence calls for the development of non-equilibrium methods for dynamical studies of the Milky Way \cite{haines2019implications,binney2018origin} and brings further difficulties for direct comparisons between cosmological simulations and our galaxy. While the time evolution of $\rho_{\odot}$ and $f(v)$ cannot be directly probed in the MW, the opposite is true for simulations, making such comparisons an attractive tool for DM studies. Figure \ref{fig:timeEvolution} shows the time evolution of both $\rho_{\odot}$ and $f(v)$ at a radius of 8 kpc from the galactic centre for Halo B and Mochima. Concerning $f(v)$ it is clear that no big changes happened in both galaxies. On the other hand, for $\rho_{\odot}$, there is a clear rise of the local density with time since the birth of the Sun but it remains within error bars. Therefore we argue that both assumptions are acceptable in the context of DM capture by the Sun, even if recent perturbations could have additional contributions on $f(v)$ and the resulting capture rate.

The dynamical history of a halo is unique, smaller halos falling into the potential well of a bigger halo get stripped of their mass. The stripped mass starts orbiting the centre of the gravitational potential and is no longer tied to the over-density in the centre of the subhalo that hosted them. The efficiency with which a potential well destroys a substructure depends on its depth and on the size and impact parameter of the substructure. Certain subhalos can get dissolve after a small number of orbits but all their mass stays in similar orbits. If enough mass is trapped in a particular set of orbits, elongated structures are formed in phase space and are observable as bumps in the velocity distributions of the DM at a certain radius. This type of bumps has been observed in cosmological simulations in the past \cite{vogelsberger2009phase} and we observe similar bump-like structures in the VDF of the Mochima simulation at 8 kpc. This bump is visible in  Figure \ref{fig:timeEvolution}. We choose to remove it from the  main distribution by tracking back in time the group of sub-halos that contribute  to the formation of the filamentary structures in phase space around $z=1.9$  and considering them as one particle \footnote{The details of this procedure will be explained in a dedicated publication in preparation that explores extensively the DM aspects of the Mochima simulation.}. After subtraction, the difference in the amplitude is due to the renormalization of the corrected distribution. In section \ref{sec:consequences} we explore the effect that this type of fluctuation in the VDF could have on the stellar capture of DM.

\subsubsection{Data extrapolation: the challenge of fitting}\label{sec:fitting}

As discussed in \citep{Ling2009,Bozorgnia:2017brl}, the galaxies obtained by hydrodynamical simulations display velocity distribution at 8 kpc from the centre that departs from classical or generalized Maxwellian shapes. In particular two problematic regions of the distribution present a pathological departure from what is observed in simulations: the population of the high velocity tail and the population around the maximum of the distribution. This features are also present in DM only simulations  (see also \cite{springel2008aquarius,Adrian-Martinez2016,Diemand2006}).
\begin{figure}[h!]
  \centering
    \begin{subfigure}[b]{0.47\linewidth}
       \includegraphics[width=\linewidth]{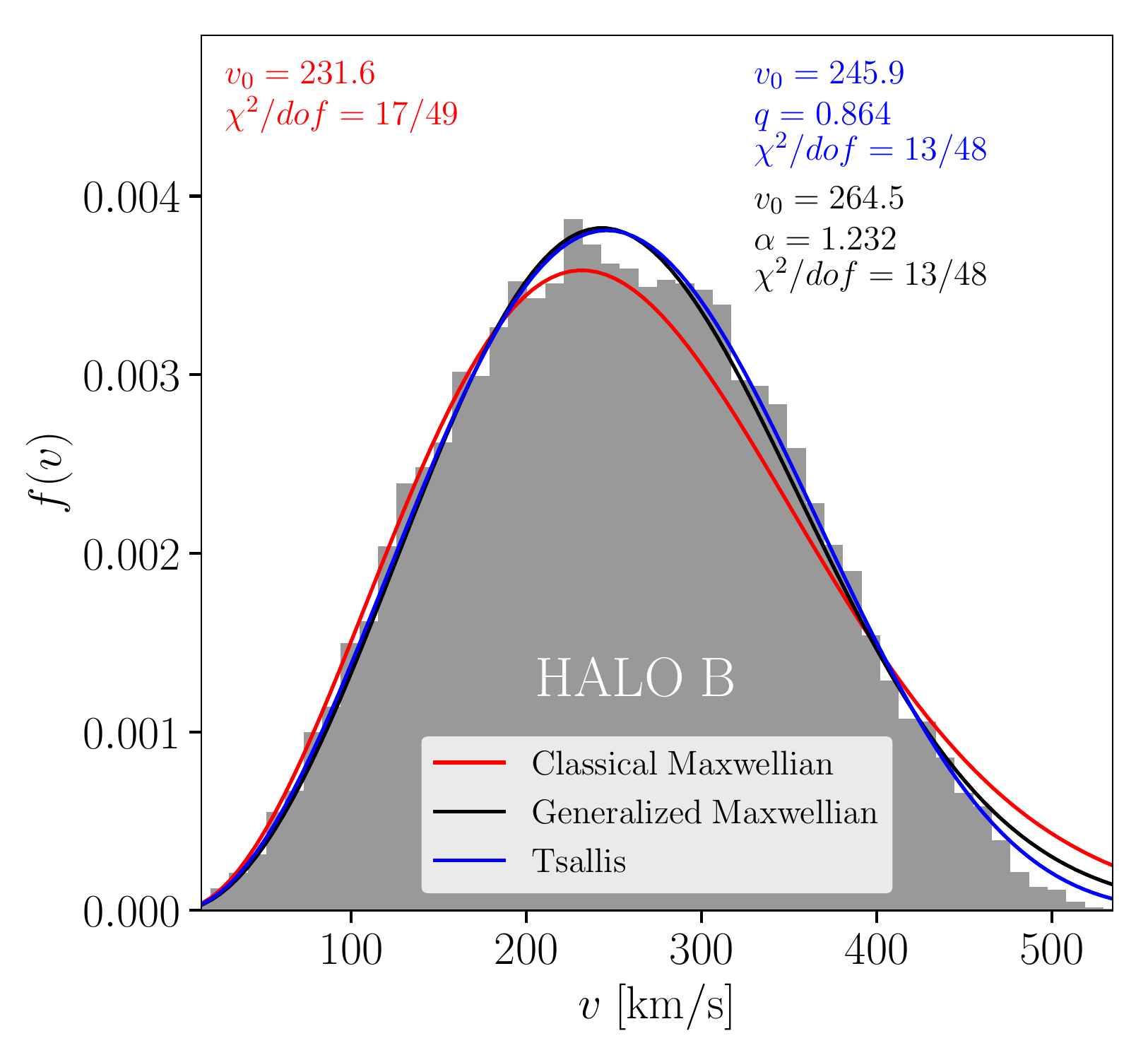}
        \caption{Halo B}
       
    \end{subfigure}
  \begin{subfigure}[b]{0.47\linewidth}
    \includegraphics[width=\linewidth]{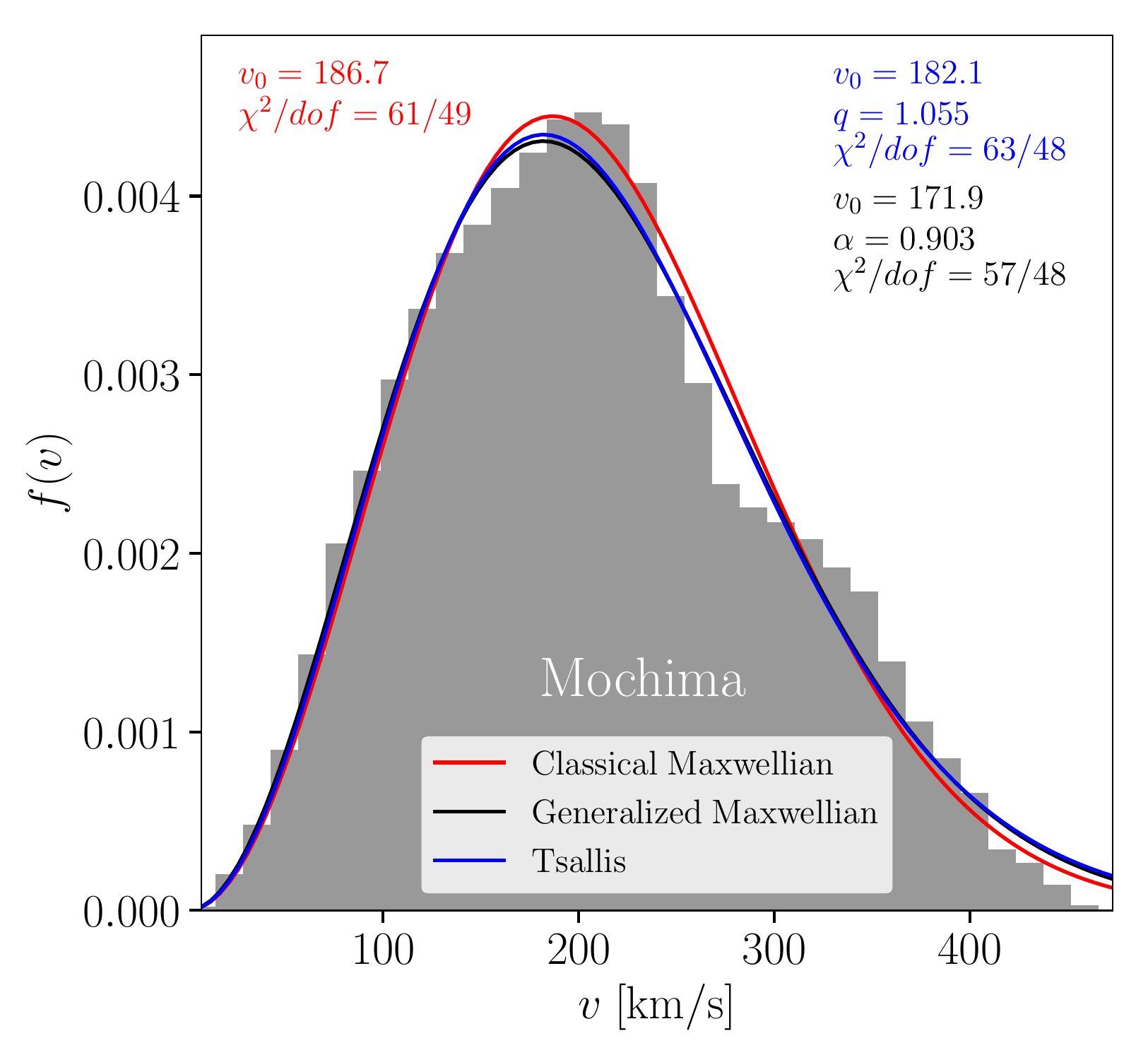}
    \caption{Mochima}
  \end{subfigure}
  \begin{subfigure}[b]{0.47\linewidth}
       \includegraphics[width=\linewidth]{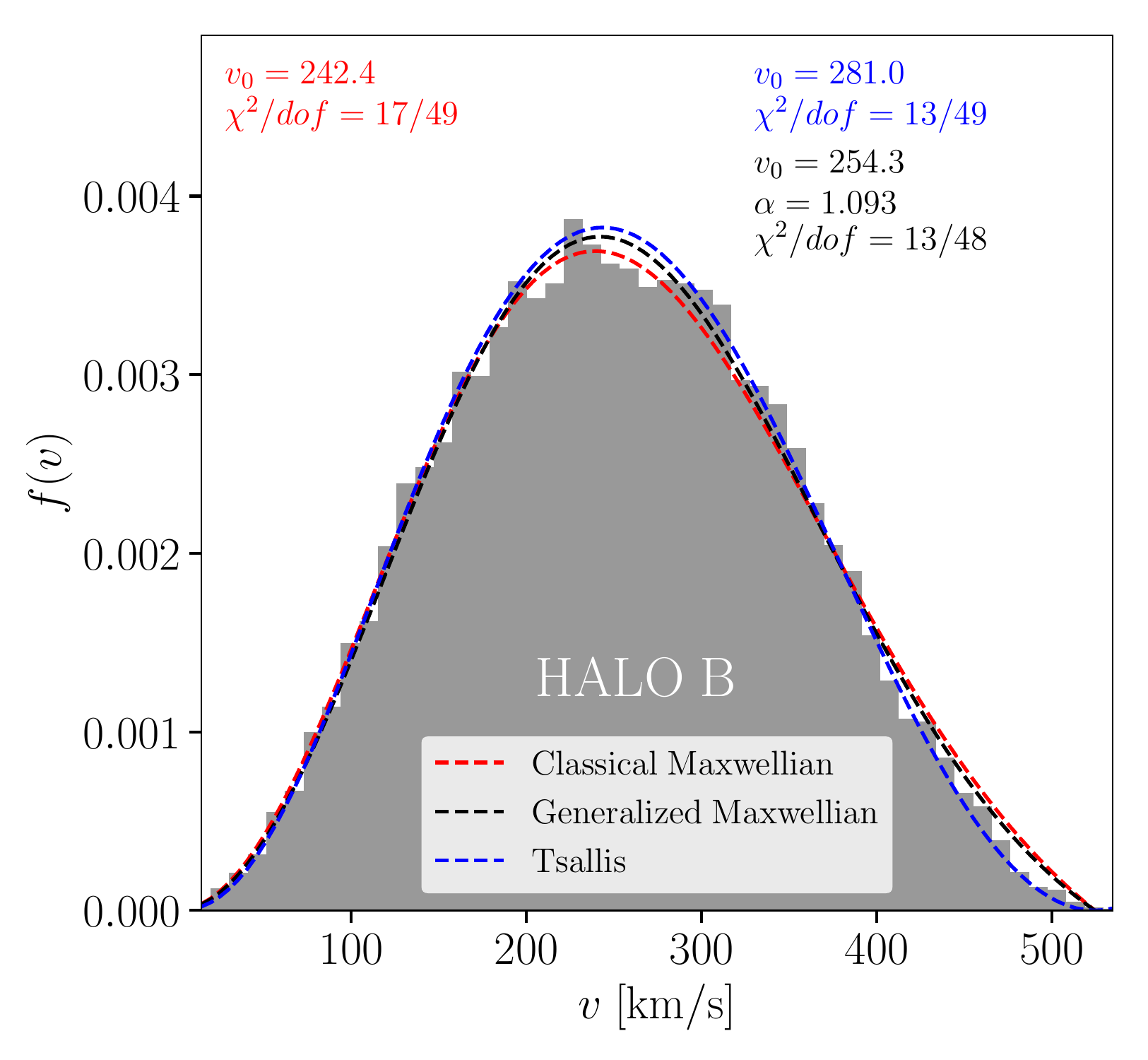}
        \caption{Halo B with $v_{esc}$}
       
    \end{subfigure}
  \begin{subfigure}[b]{0.47\linewidth}
    \includegraphics[width=\linewidth]{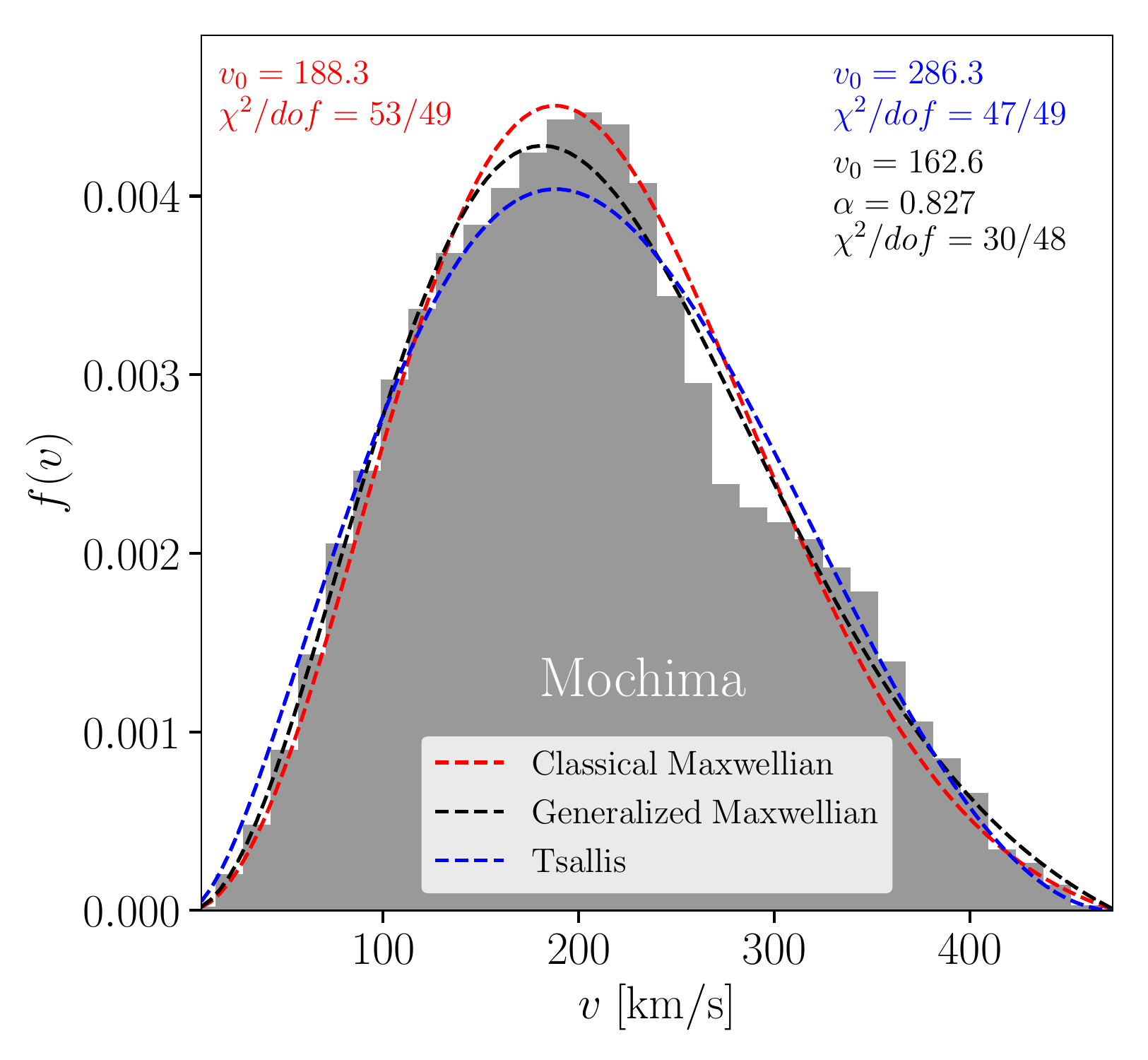}
    \caption{Mochima with $v_{esc}$}
  \end{subfigure}
  \caption{The velocity distribution of DM in a shell of 1 kpc thickness around 8 kpc, fitted with a classical Maxwellian (red), generalized Maxwellian (black) and Tsallis distribution (blue) without (top) and with (bottom) the cut at $v_{esc}$. For Mochima the corrected distribution is shown.}
  \label{fig:fdvFits}
\end{figure}
Figure \ref{fig:fdvFits} (upper panels) shows the results of fitting a classical Maxwellian over the VDF observed in both simulations. The resulting distribution overpredicts the high velocity population while underpredicting the population around the mean. This problem becomes less dramatic when considering the generalized Maxwellian distribution (equation \ref{eq:GenMaxwellian}) and the disagreement is further decreased for the Tsallis distribution. Nevertheless, both regions, the peak and the tail, remain overpredicted by both functionals \cite{Ling2009}.

\subsection{Contributions from dynamics}\label{sec:Dynamics}

  To solve the high velocity over-population in the fits we propose an additional procedure using the value of the gravitational potential at the border between the galaxy and its closest most massive neighbour, $r_{max}$ to define the local escape speed i.e the speed needed to escape the halo at a radius $r$:
\begin{equation}
v_{esc}(r) = \sqrt{2(\Phi(r_{max})-\Phi(r))}\doteq \sqrt{2 \Psi(r)}
\end{equation} 

\noindent where $\Phi(r)$ ($\Psi(r)$) is the (relative) gravitational potential (see \cite{Lacroix2018, LacroixNunez} for details).
 
\noindent The Maxwellian and the generalized Maxwellian distribution can be corrected by subtracting the value of the function at the local escape speed. For the Tsallis distribution it is enough to redefine the $q$ parameter as $q=1-(v_0^2/v_{esc}^2)$. The resulting distributions are shown in the lower panels of Figure \ref{fig:fdvFits}. A better agreement in the reproduction of the behaviour around the peak is observed by ensuring that no WIMP in the fitted distribution has a larger velocity than the local escape speed.

The escape velocity at 8 kpc for HaloB and Mochima is $v_{esc}=535$ km/s and $v_{esc}=484$ km/s respectively. The inclusion of $v_{esc}$ using the gravitational potential to approach the problem of the high velocity tail population also reduces the disagreement between the fitted distribution and the observed data, as shown by the $\chi^2$ values indicated in Figure \ref{fig:fdvFits}. Nevertheless, a direct extrapolation from a simulation to the real galaxy holds very little warranty to represent reality.

To go a step further, we consider now a dynamical approach. The full phase space Distribution Function (DF) of a non-collisional DM distribution inside a self gravitating system can be built from its structural features. Following the Jeans theorem, for a spherically symmetric system any steady state solution of the collision-less Boltzmann equation can also be written as a function of isolated integrals of motion, the energy $\mathcal{E}$ and the modulus of the angular momentum $\mathcal{L}$ (see e.g \cite{Binney,Ollongren1962,Lacroix2018}). In addition, if the system is assumed to be isotropic, this picture can be further simplified as the angular momentum becomes irrelevant and the resulting DF only depends on the energy $f(\vec{v},\vec{r})=f(\mathcal{E})$. Using the Eddington formula \cite{Eddington1916} the DF is obtained from the mass distribution by:

\begin{equation}
f(\mathcal{E})= \frac{1}{\sqrt{8} \pi^2} \frac{\text{d}}{\text{d}\mathcal{E}}\int_0^{\mathcal{E}}\frac{\text{d}\Psi}{\sqrt{\mathcal{E}-\Psi}}\frac{\text{d}\rho}{\text{d}\Psi} \ .
\end{equation}

\noindent Here $\Psi$ is the relative gravitational potential and $\rho$ the matter density. The velocity and speed distributions can then be obtained.

Cosmological simulations may be used as a testing case for this type of methods providing that the assumption of spherical symmetry is applicable for the considered object \cite{LacroixNunez}. Note that this assumption is also used in the Maxwellian approach, where the halo is also required to have $\rho(r)\propto r^{-2}$ while in the Eddington approach the density profile is the input information. The VDF obtained by the Eddington approach when compared to the Maxwellian approach yields a prediction with more consistent motivations leading to better agreement in the resulting moments of the velocity over the full halo \cite{LacroixNunez}.

Along this line of thought we have chosen to go a step further and use the Eddington inversion method to derive the DM velocity distribution at a radius of 8 kpc. To point out the variability of $f(v)$ we apply this method on the two simulations and the mass model of the Milky Way derived in \cite{Mcmillan2016}.

\begin{figure}[h!]
  \centering
  \begin{subfigure}[b]{0.47\linewidth}
    \includegraphics[width=\linewidth]{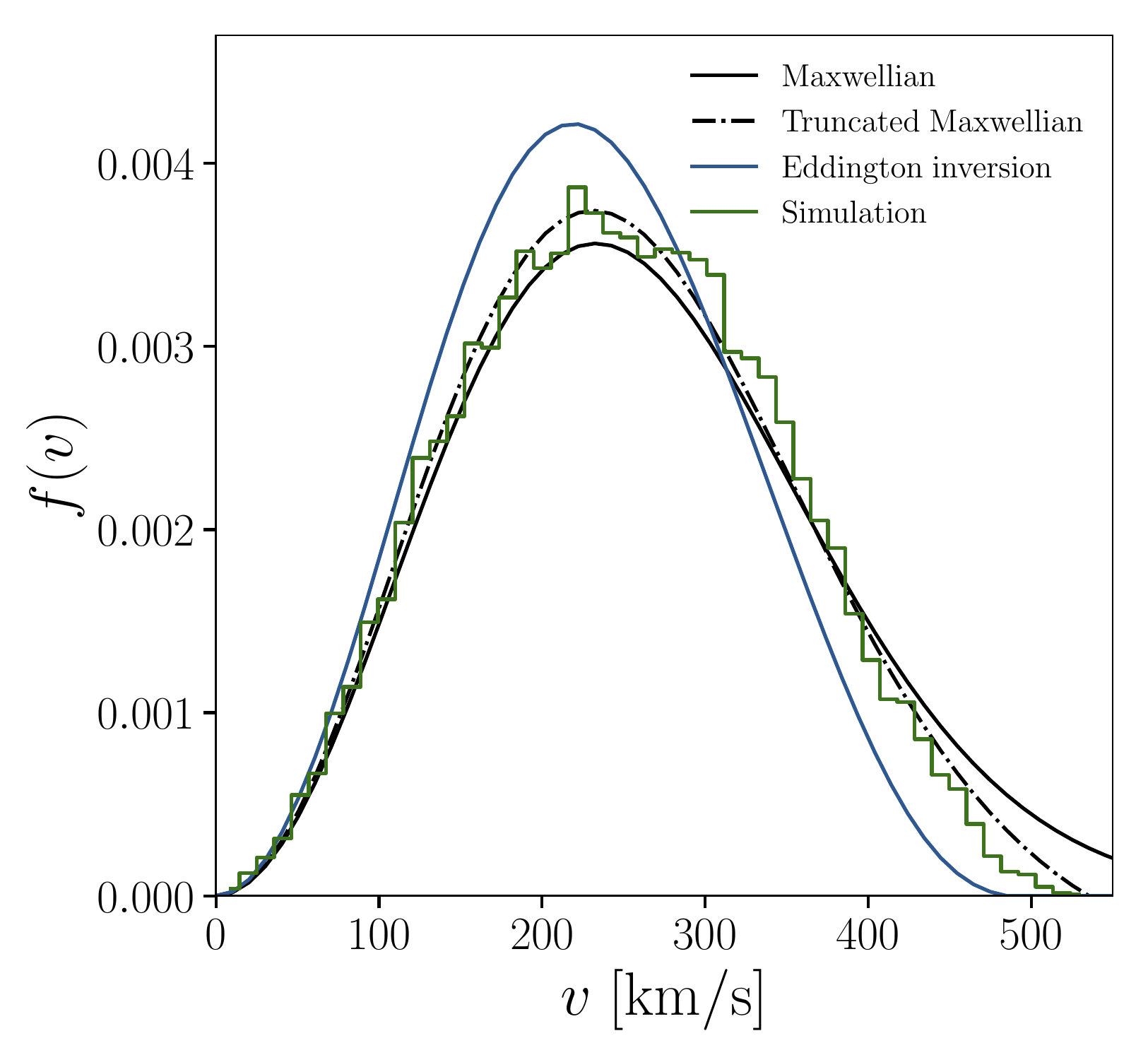}
    \caption{Halo B, galactic reference frame}
  \end{subfigure}
  \begin{subfigure}[b]{0.47\linewidth}
    \includegraphics[width=\linewidth]{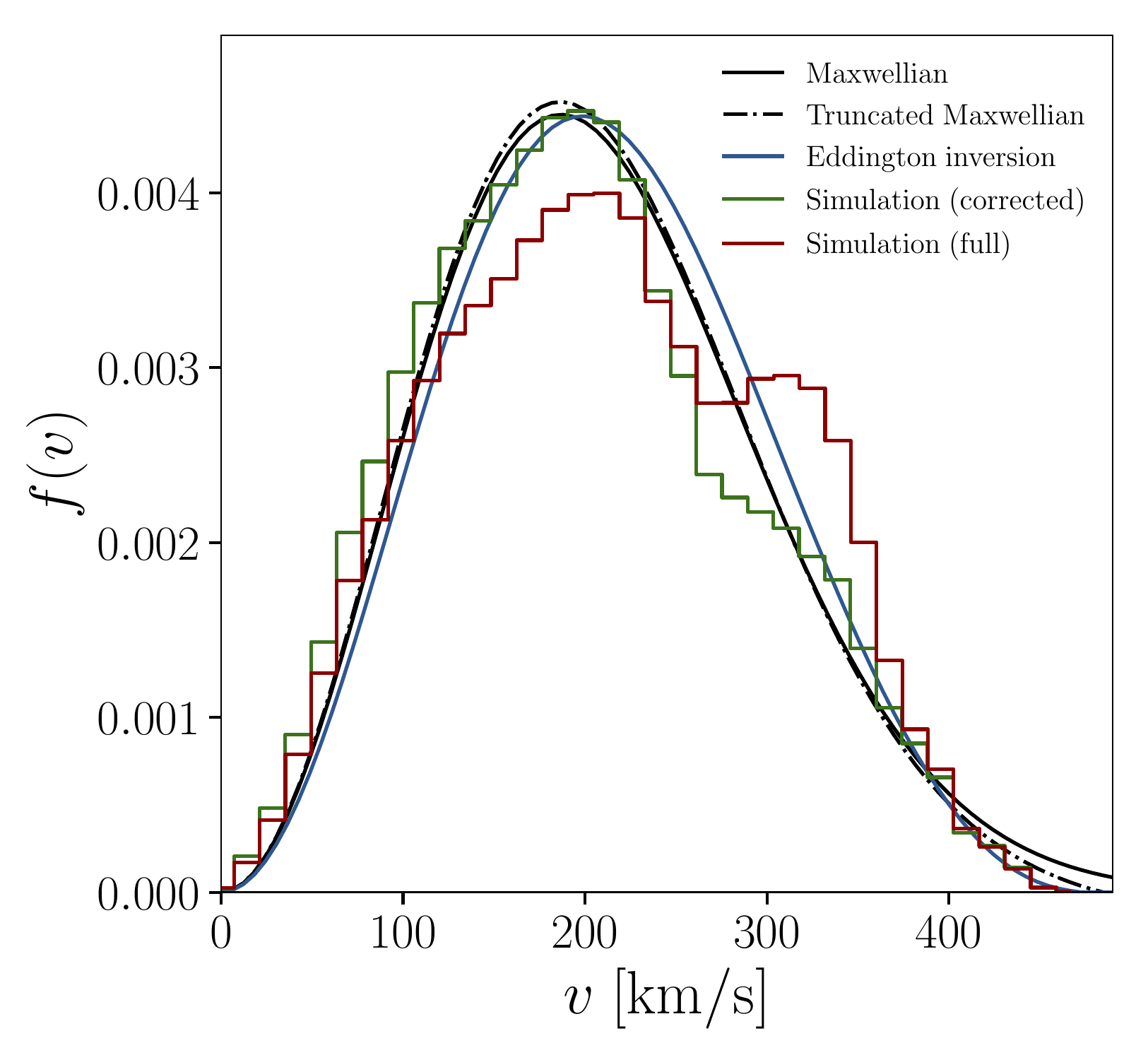}
    \caption{Mochima, galactic reference frame}
  \end{subfigure}
  \label{fig:coffee}
  \begin{subfigure}[b]{0.47\linewidth}
    \includegraphics[width=\linewidth]{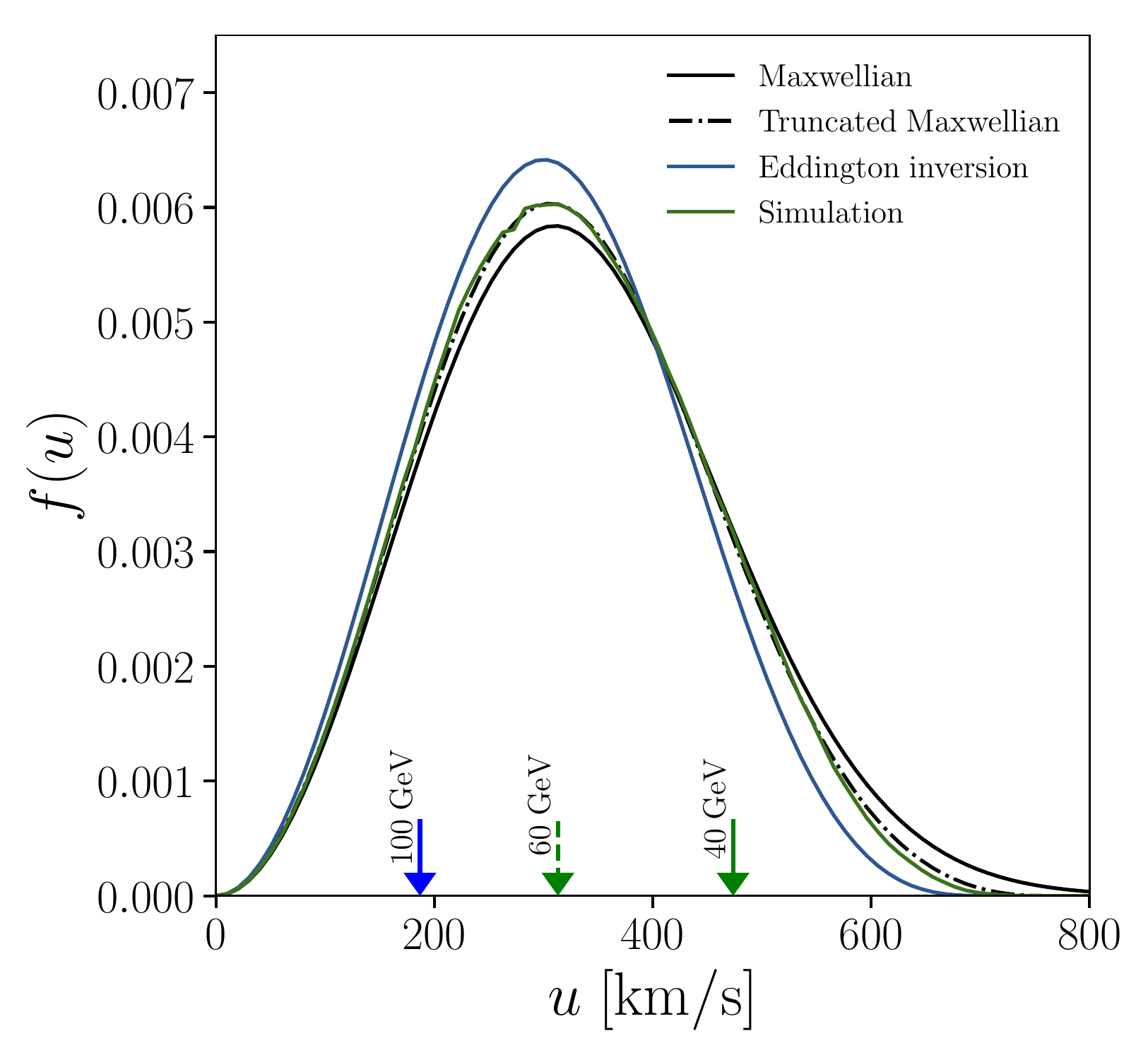}
    \caption{Halo B, solar reference frame}
  \end{subfigure}
  \begin{subfigure}[b]{0.47\linewidth}
       \includegraphics[width=\linewidth]{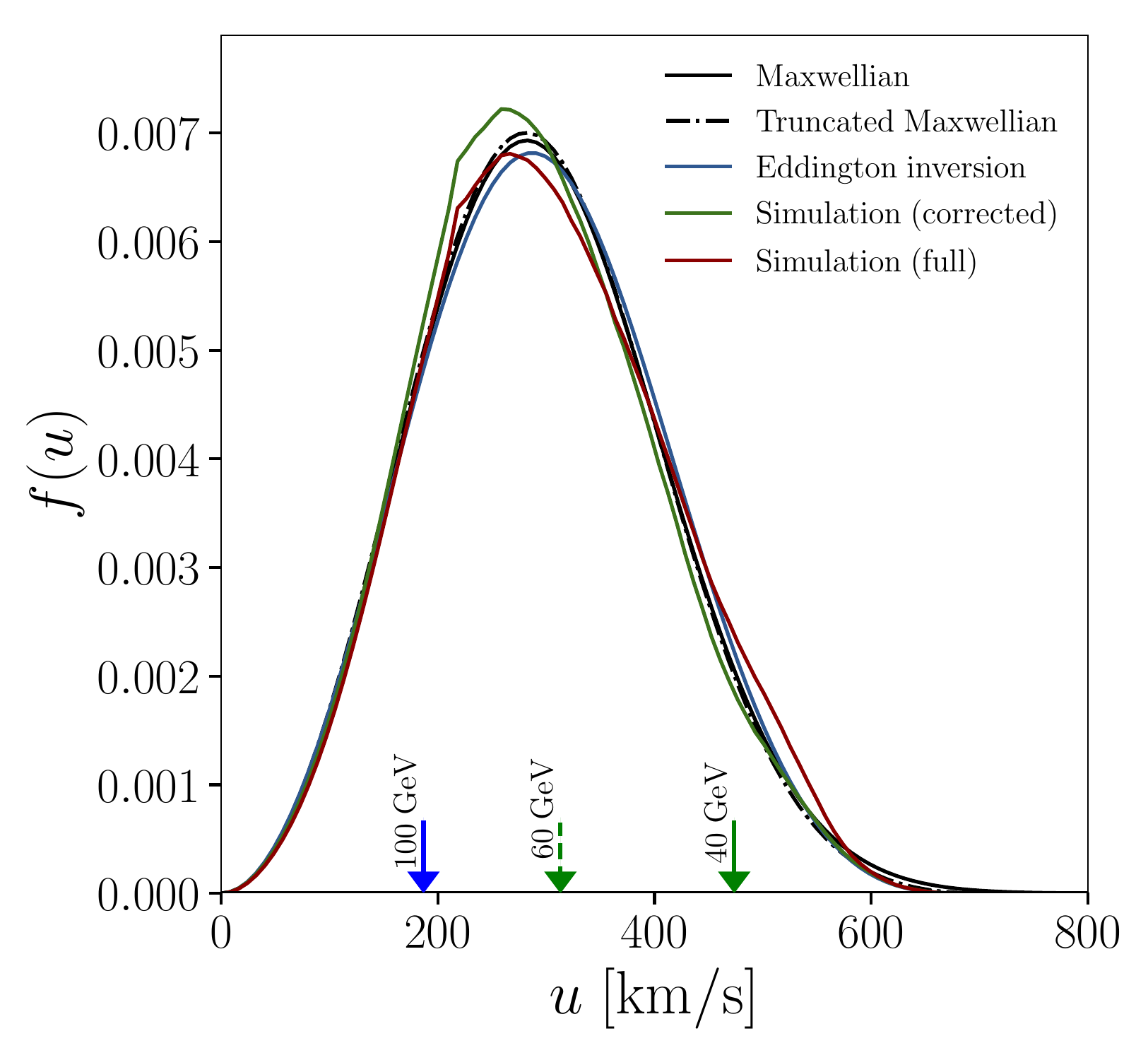}
        \caption{Mochima, solar reference frame}
    \end{subfigure}
 
  \caption{Velocity distributions of DM in the galactic (upper) and solar reference frame (bottom). The predicted velocity distributions from the Eddington method (blue) and  Maxwellian approach (black) are compared with the velocity distributions obtained in both simulations (green and red). The arrows indicate the velocities for which the capture cutoff occurs for $m_{\chi}$ equals to 100 GeV, 60 GeV and 40 GeV respectively from left to right.}
  \label{fig:fdv_And_fdu}
\end{figure}

Figure \ref{fig:fdv_And_fdu} shows for both simulations, Halo B and Mochima, the velocity distributions in the galactic frame of reference and in the solar frame of reference. The prediction of the velocity distribution using the Maxwellian approach is done by calculating the mean of the distribution as the circular velocity. This is computed from the contained mass $m(r)$ at a radius $r$ as: 
\begin{equation}
  v_c(r) = \sqrt{\frac{G m(r)}{r}}
  \label{eq:vcirc}
 \end{equation}

\noindent where $G$ is the gravitational constant. The values obtained at 8 kpc for Halo B and Mochima are $v_c=233$ km/s and $v_c=187$ km/s respectively. The resulting distributions from this approach are shown in Figure \ref{fig:fdv_And_fdu}. The mean velocity of the observed populations of DM particles is higher than those predicted by both approaches but the difference is smaller when compared to the Eddington method results. The infinite tail of the Maxwellian distribution does not describe the behaviour of the observed $v_{esc}$ while a cutoff is present in the predictions from the Eddington approach without any extra correction. We observe a better agreement with the predictions of the mean and the escape velocity for the Eddington approach for the case of the Mochima halo but not for Halo B. The maximum value of the population is systematically higher in the predictions than for the simulations which, as explained in detail in section \ref{sec:consequences}, induces a boost of the dark matter capture rate for $m_{\chi}$ that have a cutoff around the mean of the $|\vec{v}|$ of the DM particles in the local distribution.

In the next section we discuss the effect of considering different velocity distribution functions on the capture process.

\section{Consequences for the capture rate}
\label{sec:consequences}

\begin{figure}[h!]
  \centering
  \begin{subfigure}[b]{0.47\linewidth}
    \includegraphics[width=\linewidth]{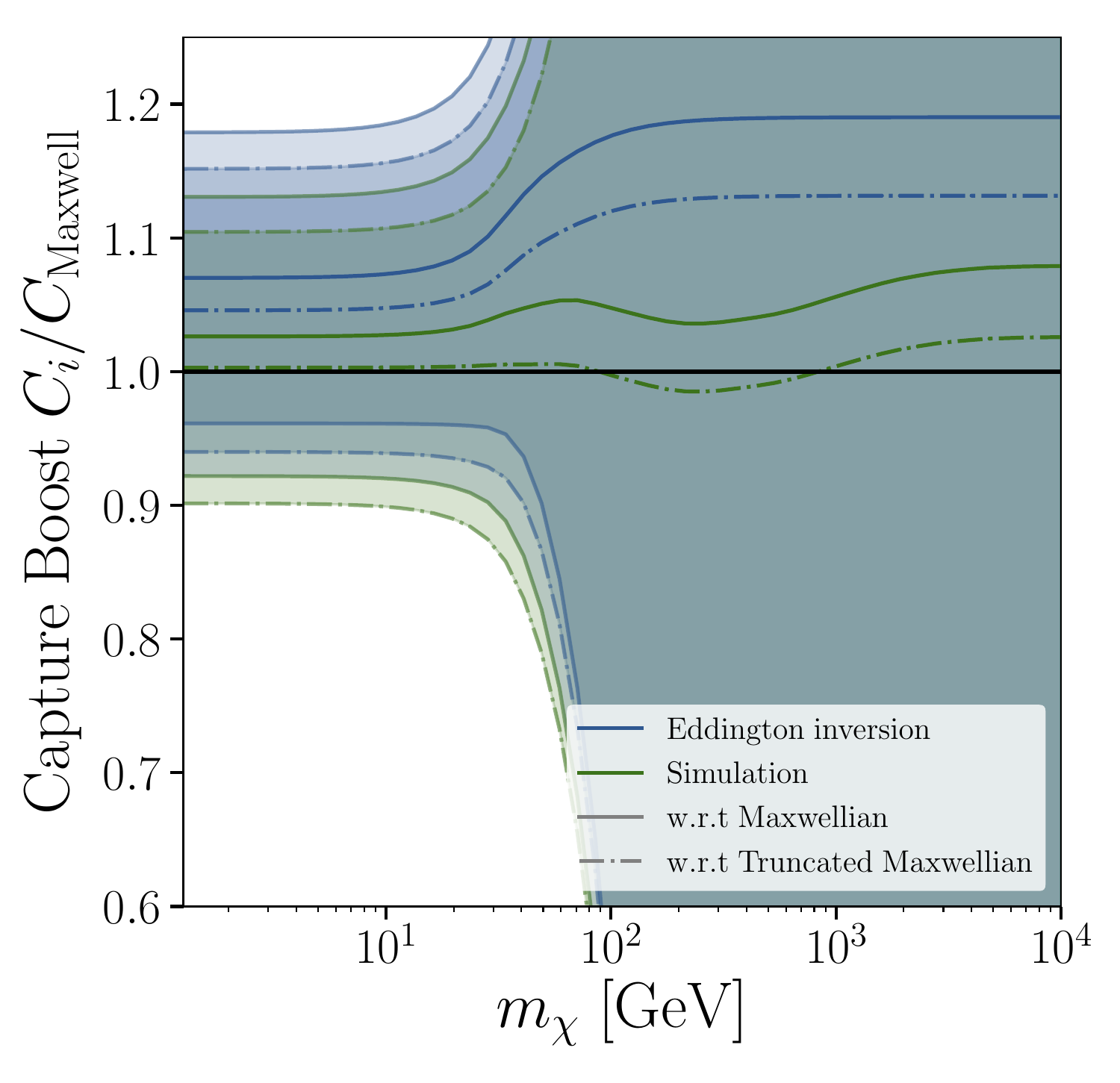}
    \caption{Halo B shell}
  \end{subfigure}
  \begin{subfigure}[b]{0.47\linewidth}
    \includegraphics[width=\linewidth]{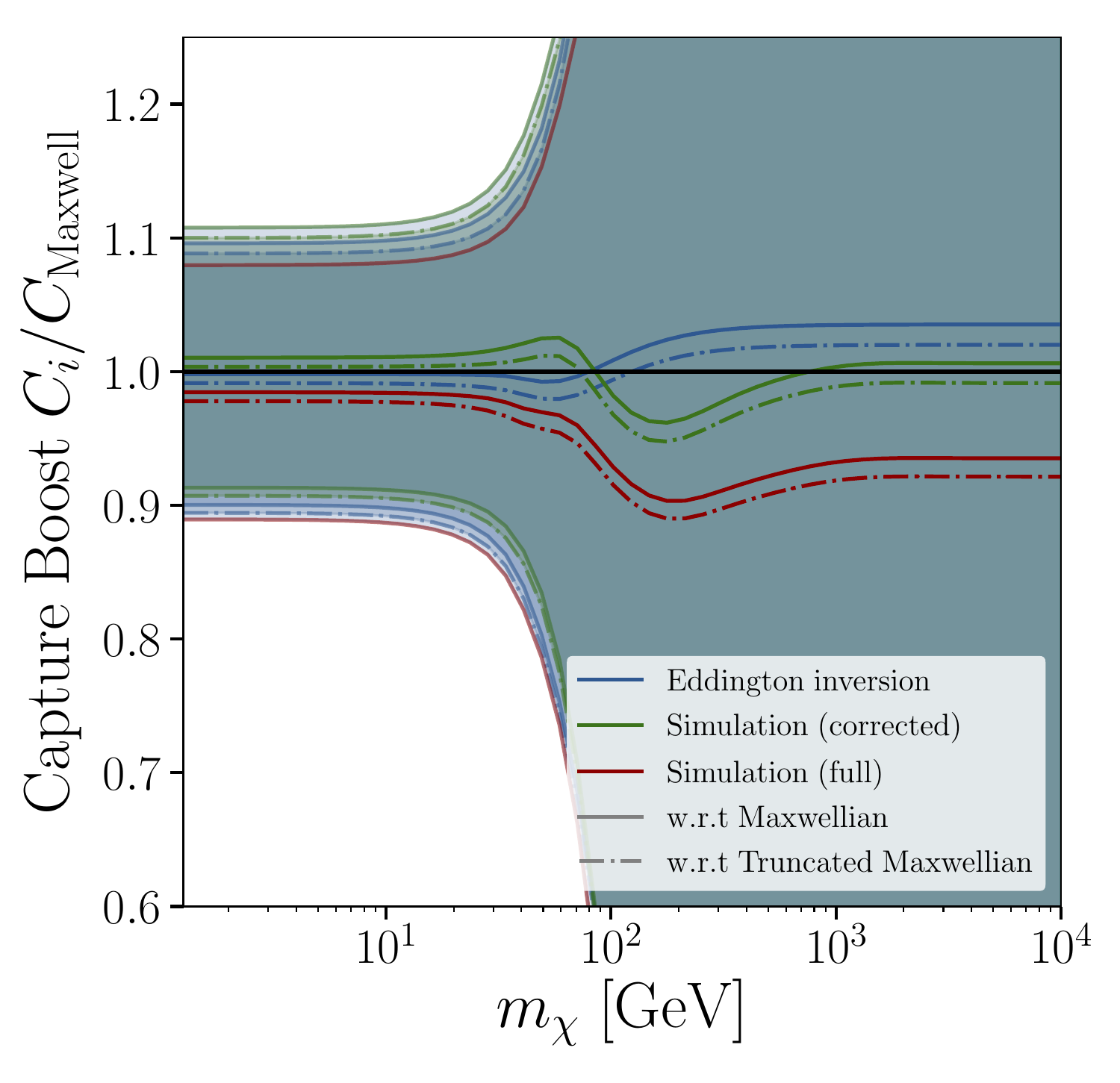}
    \caption{Mochima shell}
  \end{subfigure}
  
   \begin{subfigure}[b]{0.47\linewidth}
    \includegraphics[width=\linewidth]{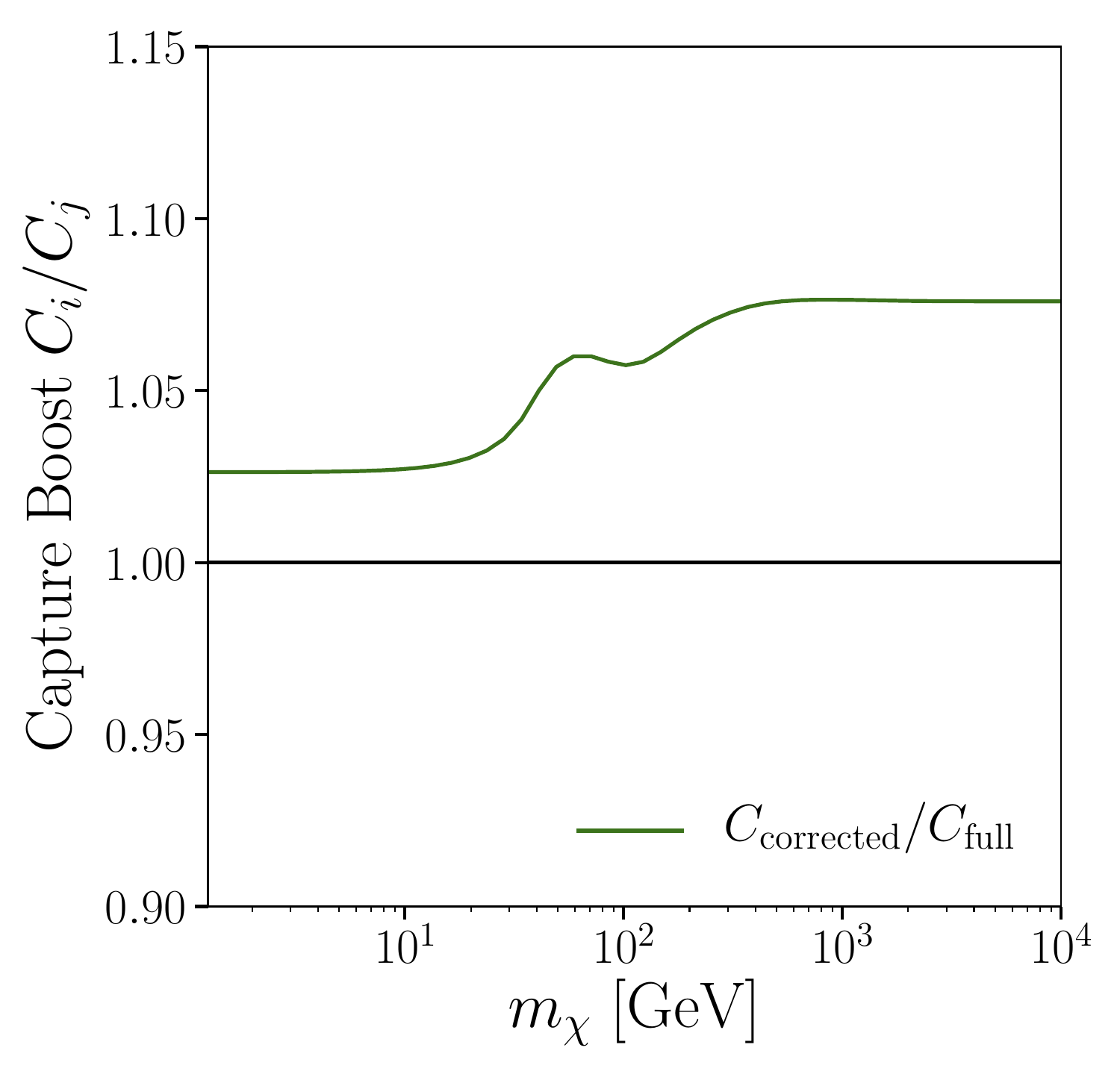}
    \caption{Mochima }
  \end{subfigure}
  \label{fig:coffee}
  \begin{subfigure}[b]{0.47\linewidth}
    \includegraphics[width=\linewidth]{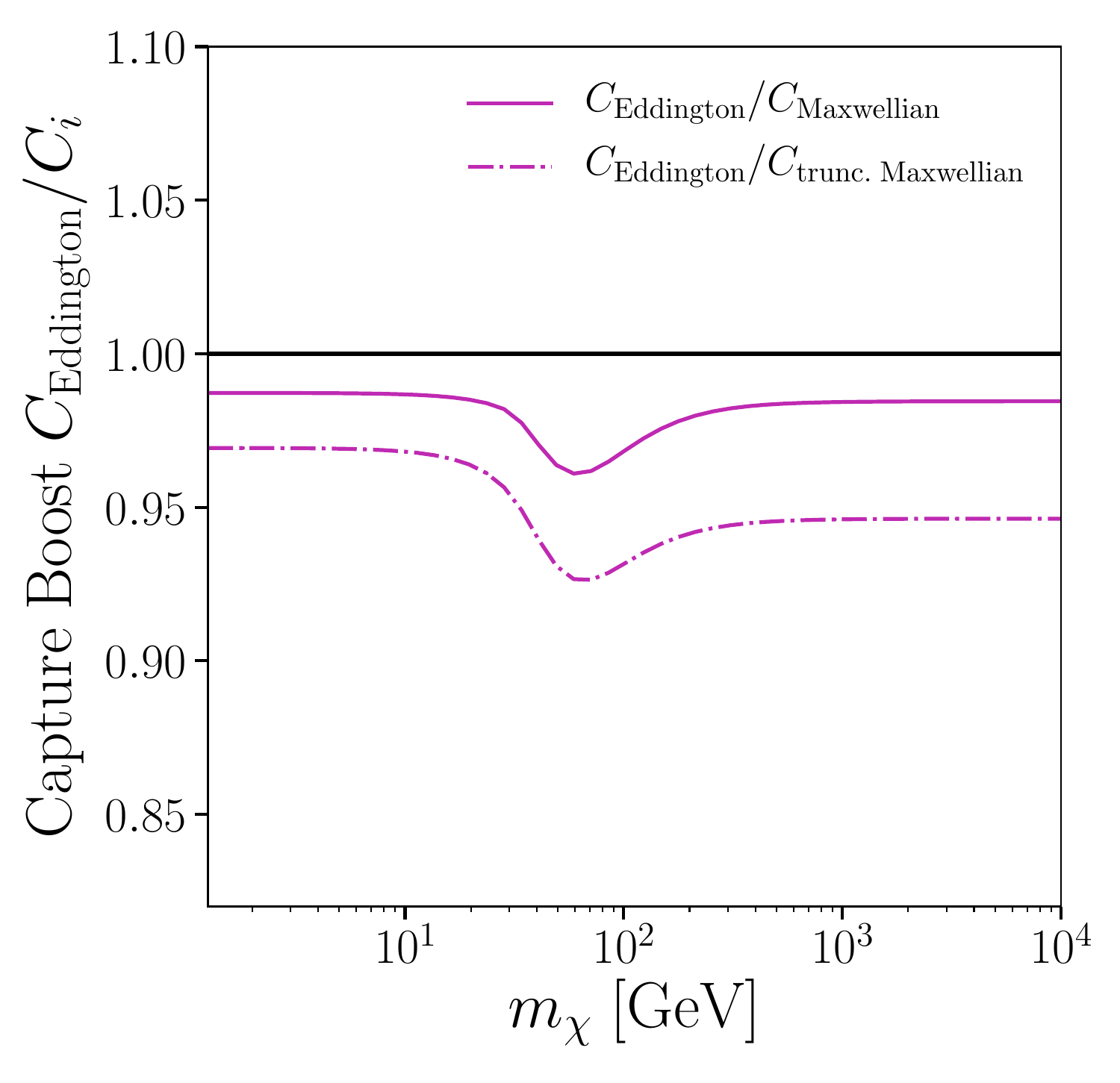}
    \caption{Milky Way mass model
    }
  \end{subfigure}

  \caption{The boost on the capture rate with respect to the Maxwellian approach (solid lines) and with respect to the truncated Maxwellian (dashed lines) for the simulated galaxies in the top panels, Halo B (a) and Mochima (b). Predictions from the Eddington method are shown in blue and the results from simulations in green, the bands shown correspond to $\sigma/10$. On panel c), the boost of the Mochima distribution without the bump population is compared to the full population. The panel d) shows the boost of the Eddington method predictions with respect to the Maxwellian and truncated Maxwellian distributions, all using parameters of the Milky Way mass model from \cite{Mcmillan2016}.  }
  \label{fig:boost}
\end{figure}

Following the previous discussion, we examine several assumptions for the dark matter velocity distribution function and evaluate the resulting impact on the capture by the Sun, illustrated in figure \ref{fig:boost}. Concerning both simulations, we consider the VDF extracted from the data (between 7 and 9 kpc from the centre of the galaxy) as well as the VDF derived from the Eddington inversion and compare them to the Maxwellian function (both truncated and untruncated). We also show the impact of the bump in $f(v)$ exhibited by the halo labelled Mochima and coming from past and recent merger events (see section \ref{sec:timeev}). Finally, we compare the capture with the VDF obtained through the Eddington approach applied on a Milky Way mass model \cite{Mcmillan2016} with regard to the SHM. We define a boost factor as $C_{i}/C_{{\rm Maxwell}}$ with $i=$ simulation data or Eddington accordingly.
In the case of the simulations, the comparison is done with respect to a capture rate computed with a Maxwellian distribution where the peak is fixed by the circular velocity following equation \ref{eq:vcirc}.
For the Milky Way mass model we use $v_{\odot}=233$ km/s from \cite{Mcmillan2016}.


Results of the capture boost are shown in figure \ref{fig:boost}. Generally we see that for heavy WIMPs ($m_{\chi}>100$ GeV), for which the capture is only sensitive to the low velocity population, the absolute value of the boost is larger than for light WIMPs, increasing or decreasing the capture rate depending on the case. We observe that the rise in the boost happens for  $m_{\chi}> 50$ GeV where the cutoff occurs at the mean of the VDF (illustrated by the arrows in figure \ref{fig:fdv_And_fdu}). The capture rate in the Eddington case has a positive boost for heavy WIMPs in both simulations because the Maxwellian distribution overpredicts the high speed population. The velocity cutoff first affects the tail of the Maxwellian distribution by comparison to the Eddington distribution as $m_{\chi}$ increases.

 The Eddington prediction for Halo B is underestimating the number of fast WIMPs and this difference results in a positive boost of $20\%$ for heavy WIMPs. In the Mochima halo the overpredicted high velocity population is less dramatic. Here there are two crossings between the  Maxwellian and the Eddington distributions, meaning that a higher population is predicted by the Eddington method  for WIMPs with $350\; \mathrm{km/s} < v< 550 $ km/s. This results in an inversion of the boost from a positive boost of $4\%$ for heavy WIMPs to a decreasing of few percents in the range of $30\ \mathrm{GeV} < m_{\chi} < 80 $ GeV.
 
 For the Mochima simulation we consider an extra case, the {\it corrected} velocity distribution, meaning that we have removed the subhalos that form the bump in the observed VDF because the functions used for the fit of the distribution cannot reproduce the features created by those subhalos. It has been argued that the capture rate is not sensitive to DM substructures due to the long time during which the capture process takes place \cite{Choi2014}. We find that in the Mochima halo the distribution belonging to the bump is present  during the full lifetime of the Sun, as shown in figure \ref{fig:timeEvolution}. To illustrate the contribution of this bump, we show on figure \ref{fig:boost} the boost between the corrected and full $f(v)$.  The capture rate is  $5\%$ less efficient by considering  the corrected distribution. This is because the DM particles that populate the structures in phase space in the Mochima halo constitute a non-negligible fraction of the local DM population. This effect highlights the importance that the halo history and substructure dynamics might have on DM detection sensitivity and the limitations of considering only smooth VDFs that predict at most some general features of the halo.

For the sake of completeness, we propose now to apply the Eddington inversion to the Milky Way mass model derived in \cite{Mcmillan2016}. The comparisons with the Maxwellian velocity distribution of the SHM are shown on figure \ref{fig:boost}. We obtain some differences between 5-10\% for the untruncated  and 5-15\% for the truncated case.
 
 Finally, if we propagate the intrinsic uncertainty of the capture rate calculation discussed in section \ref{sec:uncertainties} to the capture boost, the resulting uncertainty band is so big that only $1/10$ of its total value is shown in figure \ref{fig:boost} upper panels. The uncertainties are particularly divergent for heavy WIMPs making predictions in those regions difficult to really trust. We emphasize  that this should be more cautiously taken into account when discussing the detection of dark matter captured by the Sun with neutrino telescopes.

\section{Summary/conclusions}\label{sec:Summary}

The detection of a neutrino signal coming from dark matter captured in the Sun is a golden search for current and future neutrino telescopes. It would be a compelling evidence  for the presence of local dark matter. In this paper we discuss the astrophysical assumptions relevant for the dark matter capture in the Sun mainly focusing on the dark matter velocity distribution in the solar neighbourhood. To highlight the related uncertainties and the consequences on the resulting capture rate, we revisit usual functions used as popular ansatz for those calculations, testing their ability to fit simulation data of two state of the art cosmological simulations of ``Milky Way like'' galaxies. For completeness and comparison, we also apply the Eddington inversion on the two simulations and on a mass model of the Milky Way. Every assumption is compared to the Maxwellian velocity distribution function commonly used in the Standard Halo Model.

Our main results are the following :
\begin{itemize}
\item[$-$] Even if the Sun could fall out of the capture/annihilation equilibrium during a fluctuation period of low capture, this is an unlikely scenario and the Sun is safely in  equilibrium for WIMPs with $m_{\chi}\lesssim $ few TeV. 
\item[$-$] The peak/hat and the tail  of the simulations are typically hardly reproduced by usual functions. Adding the escape velocity in the fits improves the consistency with the tail of the distribution.
\item[$-$] The VDFs derived through the Eddington approach bring additional information on the possible distributions that can be assumed and have a better agreement with simulations data compared to the standard Maxwellian VDF.
\item[$-$] We confirm that merger history of the halo can leave specific features in the velocity distribution function that cannot be taken into account by usual functions (at least considering a single VDF).
\item[$-$] The level of variability on the capture rate can reach up to 20\% depending on the assumptions made for the dark matter velocity distribution.
\item[$-$] The highlighting of the intrinsic errors  by evaluating the variance of the capture rate leads to dramatic and disturbing uncertainties, especially for WIMP masses larger than 30 GeV, opening questions on the reliability of predictions in this range of mass.
\end{itemize}
  
Meanwhile, the sustained efforts to upgrade experimental sensitivities of local dark matter searches, the {\it Gaia} survey opens a precision astrometry era and should bring strong  improvements on our knowledge of the dark matter distribution properties reducing the speculations about the relevant direct and indirect detection aspects.

\acknowledgments

This work has been carried out thanks to the support of the OCEVU Labex (ANR-11-LABX-0060) and the
A*MIDEX project (ANR-11-IDEX-0001-02) funded by the Investissements d'Avenir French government
program managed by the ANR.

This work was granted access to the HPC resources of Aix-Marseille Universit\'e financed by the project Equip@Meso (ANR-10-EQPX-29-01) of the program  ``Investissements d'Avenir'' supervised by the Agence Nationale de la Recherche.


\bibliography{Capture_ANC_VB_EN} 

\bibliographystyle{JHEP}








\end{document}